\documentclass{aa}
\usepackage{amsmath}
\usepackage{txfonts}
\usepackage{graphicx}
\usepackage{natbib}
\usepackage{amsfonts}
\usepackage{amssymb}
\usepackage{hyperref}
\usepackage{longtable}
\usepackage{color}

\bibpunct{(}{)}{;}{a}{}{,} 

\renewcommand{\hat}[1]{\oldhat{\mathbf{#1}}}

\begin{document}
\title{Study of KIC 8561221 observed by  {\emph{Kepler}}:\\ an early red giant showing depressed dipolar modes}
\author{R.~A.~Garc\'\i a\inst{1,2} \and 
 F. P\'erez Hern\'andez \inst{3,4}\and
O.~Benomar\inst{5}\and
 V.~Silva~Aguirre \inst{6,2} \and
J.~Ballot \inst{7,8} \and
G.~R. Davies \inst{9,1} \and
G. Do\u{g}an\inst{10,2}\and
D.~Stello \inst{5} \and
 J.~Christensen-Dalsgaard \inst{6,2} \and
G. Houdek \inst{6,2}\and
F. Ligni\`eres \inst{7,8,2}\and
 S.~Mathur\inst{10,11,2} \and
M. Takata \inst{12,2}\and
 T.~Ceillier \inst{1} \and 
 W.~J. Chaplin\inst{9,2}\and
S. Mathis\inst{1}\and
 B.~Mosser \inst{13} \and
 R.~M. Ouazzani\inst{13}\and
 M.~H. Pinsonneault\inst{14,2}\and
 D.~R.~Reese \inst{15,2}\and
 C.~R\'egulo \inst{3,4} \and
 D.~Salabert \inst{1} \and
 M.~J.~Thompson\inst{10}\and
 J.~L. van Saders\inst{14}\and
  C.~Neiner\inst{13}\and
 J. De Ridder\inst{16}
 }
 \institute{Laboratoire AIM, CEA/DSM -- CNRS - Univ. Paris Diderot -- IRFU/SAp, Centre de Saclay, 91191 Gif-sur-Yvette Cedex, France
 \and Kavli Institute for Theoretical Physics, University of California, Santa Barbara, CA 93106-4030, USA
 \and Ins\-ti\-tu\-to de Astrof\'\i sica de Canarias, 38205, La Laguna, Tenerife, Spain
 \and Uni\-ver\-si\-dad de La Laguna, Dpto de Astrof\'isica, 38206, Tenerife, Spain
\and Sydney Institute for Astronomy, School of Physics, University of Sydney, NSW 2006, Australia
\and Stellar Astrophysics Centre, Dpt. of Physics and Astronomy, Aarhus University, Ny Munkegade 120, DK-8000 Aarhus, Denmark
\and CNRS, Institut de Recherche en Astrophysique et Plan\'etologie, 14 avenue Edouard Belin, 31400 Toulouse, France
\and Universit\'e de Toulouse, UPS-OMP, IRAP, 31400 Toulouse, France
\and School of Physics and Astronomy, University of Birmingham, Edgbaston, Birmingham B15 2TT, UK
\and High Altitude Observatory, National Center for Atmospheric Research, P.O. Box 3000, Boulder, CO 80307, USA
\and Space Science Institute, 4750 Walnut Street, Suite 205, Boulder, Colorado 80301, USA
\and Department of Astronomy, School of Science, The University of Tokyo, 7-3-1 Hongo, Bunkyo-ku, Tokyo 113-0033, Japan
\and LESIA, Observatoire de Paris, CNRS UMR 8109, UPMC, Universit\'e Paris Diderot, 5 place Jules Janssen, 92190 Meudon, France
\and Astronomy Department, Ohio State University, Columbus, Ohio 43210, USA
\and Institut d'Astrophysique et G{\'e}ophysique de l'Universit{\'e} de Li{\`e}ge, All{\'e}e du 6 Ao\^ut 17, 4000 Li{\`e}ge, Belgium
\and Instituut voor Sterrenkunde, KU Leuven, Celestijnenlaan 200D, B-3001 Leuven, Belgium
}
\date{Received 01 September 2013/ Accepted }
\abstract{The continuous high-precision photometric observations provided by the CoRoT and \emph{Kepler} space missions have allowed us to better understand the structure and dynamics of red giants using asteroseismic techniques. A small fraction of these stars shows dipole modes with unexpectedly low amplitudes. The reduction in amplitude is more pronounced for stars with higher frequency of maximum power, $\nu_{\rm{max}}$.}
{In this work we want to characterize KIC 8561221 in order to confirm that it is currently the least evolved star among this peculiar subset and to discuss several hypotheses that could help explain the reduction of the dipole mode amplitudes.}
{We  used \emph{Kepler} short- and long-cadence data combined with spectroscopic observations to infer the stellar structure and dynamics of KIC~8561221. We  then discussed different scenarios that could contribute to the reduction of the dipole amplitudes such as a fast rotating interior or the effect of a magnetic field on the properties of the modes. We  also performed a detailed study of the inertia and damping of the modes.}
{We have been able to characterize 37 oscillations modes, in particular, a few dipole modes above $\nu_{\rm{max}}$ that exhibit nearly normal amplitudes. The frequencies of all the measured modes were used to determine the global properties and the internal structure of the star. We have inferred a surface rotation period of $\sim$ 91 days and uncovered the existence of a variation in the surface magnetic activity during the last 4 years. The analysis of the convective background did not reveal any difference compared to ``normal'' red giants. As expected, the internal regions of the star probed by the $\ell=$ 2 and 3 modes spin 4 to 8 times faster than the surface.}
{With our grid of standard models we are able to properly fit the observed frequencies. Our model calculation of mode inertia and damping  give no explanation for the depressed dipole modes.  A fast rotating core is also ruled out as a possible explanation. Finally, we do not have any observational evidence of the presence of a strong deep magnetic field inside the star.}
\authorrunning{R.A. Garc\'\i a et al.}
\titlerunning{Study of KIC~8561221}
\maketitle

\begin{figure*}[!ht]
\centering
\includegraphics[scale=0.71]{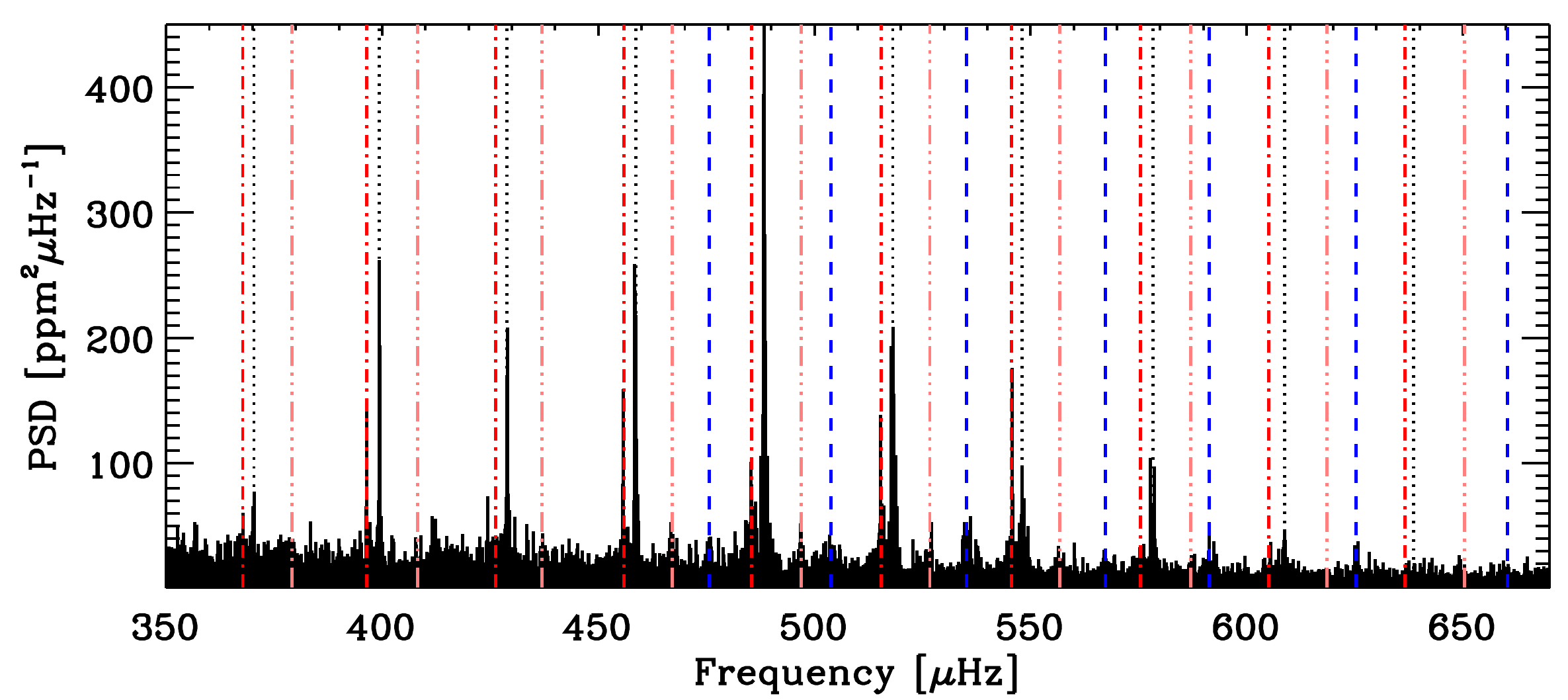}
\caption{Zoom of the Power density spectrum around $\nu_{max}$. Vertical lines indicate the central frequencies obtained in the analysis described in Sect.~\ref{fitting} and Table~\ref{tbl-2} for modes $\ell$=0, 1, 2, and 3, respectively black dot, blue dashed, red dot dashed, and pink triple-dot dashed lines. 
}
\label{fftzoom}
\end{figure*}

\section{Introduction}
Turbulent motions in the outer convective envelope of  ``solar-like pulsating stars'' excite oscillation modes that propagate inside the stars probing their interiors \citep[e.g.][]{1977ApJ...212..243G,1988ApJ...326..462G,2001A&A...370..136S}. The first indications of solar-like oscillations in red giants were obtained from ground-based observations in the K-giant Arcturus \citep[][]{1999ASPC..185..187M}  and in the G7-giant star $\xi$ Hya \citep[e.g.][]{2002A&A...394L...5F}. With the development of high-precision space photometry, the field moved rapidly from  the study of a few red giants  \citep[e.g.][]{2000ApJ...532L.133B,2007A&A...468.1033B} to multi-month continuous observations --provided by CoRoT \citep{2006cosp...36.3749B} and {\it Kepler} \citep{2010Sci...327..977B}-- of many field red giants \citep[e.g.][]{2009Natur.459..398D,2009A&A...506..465H,2010A&A...517A..22M,2010ApJ...713L.176B,2010ApJ...723.1607H}, as well as a few hundred stars belonging to clusters  \citep[e.g.][]{2010AN....331..985S,
2011ApJ...729L..10B}. These data provide strong constraints on these stars. In particular, the detection of mixed modes \citep{2011Sci...332..205B} opened a new window on the understanding of the properties of their cores \citep{2011Natur.471..608B,2011A&A...532A..86M}, including their dynamics \citep{2012A&A...548A..10M,2012ApJ...756...19D,2012Natur.481...55B}. 

In 2011, Garc\'\i a et al. showed at the $4^{th}$ \emph{Kepler} Asteroseismic Science Consortium (KASC-4) meeting, two red-giant stars observed by {\it Kepler} in which the amplitudes of the dipole modes were unexpectedly small. The following ensemble analysis by \citet{2012A&A...537A..30M} on 800 red giants  observed by {\it Kepler} \citep[e.g.][]{2010ApJ...723.1607H,2011A&A...525A.131H} showed that in about 5$\%$ of these stars the dipole modes had lower visibilities than expected from theory \citep[see Fig.~11 of ][]{2012A&A...537A..30M}. Moreover, these stars followed a relation with $\nu_{\rm{max}}$ where higher $\nu_{\rm{max}}$ stars showed lower dipole mode visibility. No clear correlation appeared with other variables such as the effective temperature or the mass, although a lower limit of the mass seems established at around 1.4 $M_{\odot}$.

To better understand the possible physical mechanisms responsible of such decrease in the dipole amplitudes, we study KIC~8561221 \footnote{Known inside KASC as Droopy}, an early red giant near the base of the red giant branch (RGB) \citep[see Fig.~3 of][]{2012A&A...540A.143M}. This is the least evolved known depressed dipole-mode star (see Fig.~\ref{fftzoom}) and, as a consequence, is the star with the strongest mode depression/suppression.  It also shows a clear transition/gradient  in the level of depression/suppression from very strong below $\nu_{\rm{max}}$ to near normal at high frequencies.  This could therefore represent the Rosetta stone for our studies of the depressed/suppressed dipole-mode phenomenon as it marks the transition where the reduction in the dipole amplitudes ``kicks in''.

KIC~8561221 was observed by the NARVAL spectrograph mounted on the 2-m Bernard-Lyot telescope at the Pic du Midi Observatory in France. Two groups analysed the spectra using different methods. \citet{2012MNRAS.423..122B} used the semi-automatic software VWA \citep{2010MNRAS.405.1907B} to retrieve  $T_{\rm eff}$ and [Fe/H] after adopting the asteroseismic $\log g$. \citet{2013MNRAS.434.1422M} used two other codes (ROTFIT and ARES+MOOG) to analyse the same spectra and reestimated $T_{\rm eff}$ and [Fe/H].
The values obtained by these different methods agree within the uncertainties and are summarised in Table~\ref{tbl-spec}. For the rest of our paper we will use the parameters obtained by \citet{2012MNRAS.423..122B}  because they took the seismic $\log g$ into account.

\begin{table}[htdp]
\caption{Spectroscopic parameters of KIC~8561221 obtained by \citet[][]{2012MNRAS.423..122B}  using the VWA method and by \citet{2013MNRAS.434.1422M}  using ROTFIT and ARES+MOOG.}
\begin{center}
\begin{tabular}{rrcc}
\hline
\hline
 KIC& $\log g$ &  $T_{\rm eff}$ (K) & [Fe/H] \\
\hline
VWA & 3.61\,$\pm$\,0.03 &  5245\,$\pm$\,60 & $-$0.06\,$\pm$\,0.06\\
ROTFIT & 3.76\,$\pm$\,0.13 & 5290\,$\pm$\,89 & $-$0.04\,$\pm$\,0.10 \\
ARES+MOOG & 3.80\,$\pm$\,0.11 & 5352\,$\pm$\,68 & $-$0.04\,$\pm$\,0.06 \\
\hline
\end{tabular}
\end{center}
\label{tbl-spec}
\end{table}%

We start in Sect.~\ref{Sec_obs} by presenting the {\it Kepler} observations used in this paper, as well as the determination of the dynamic (convection, rotation and magnetic activity) and seismic properties of the star. With this set of observables and the spectroscopic constraints from the VWA method, we compute several stellar models in Sect.~\ref{Sec_Model} using different methodologies. Finally, in Sect.~\ref{Sec_disc}, we discuss different physical mechanisms that  could, potentially, contribute to the reduction of the dipole-mode amplitudes
\section{Observations and data analysis}
\label{Sec_obs}
In this paper we have used data obtained by the NASA {\it Kepler} mission \citep{2010Sci...327..977B}. The photometric variations of most of the stars studied by {\it Kepler} are measured with a long-cadence (LC) sampling rate of 29.4244 minutes.  
In addition, {\it Kepler} also has up to 512 targets observed with a short cadence (SC) of 58.85 s \citep{2010ApJ...713L.160G} particularly aimed at detailed characterisation of planet hosting stars \citep[e.g.][]{2012ApJ...746..123H,2013ApJ...767..127H}. The SC time series are required to detect p-mode oscillations above the Nyquist frequency of the LC light curves at $\sim 283$ $\mu$Hz, which include main-sequence, subgiant, and early red-giant stars. Therefore, 829.5 days of SC data from quarters Q5 to Q13 --starting on March 20, 2010 and ending on October 3, 2012-- have been used to study the oscillation properties of KIC~8561221. 

\begin{figure*}[!ht]
\centering
\includegraphics[scale=0.75]{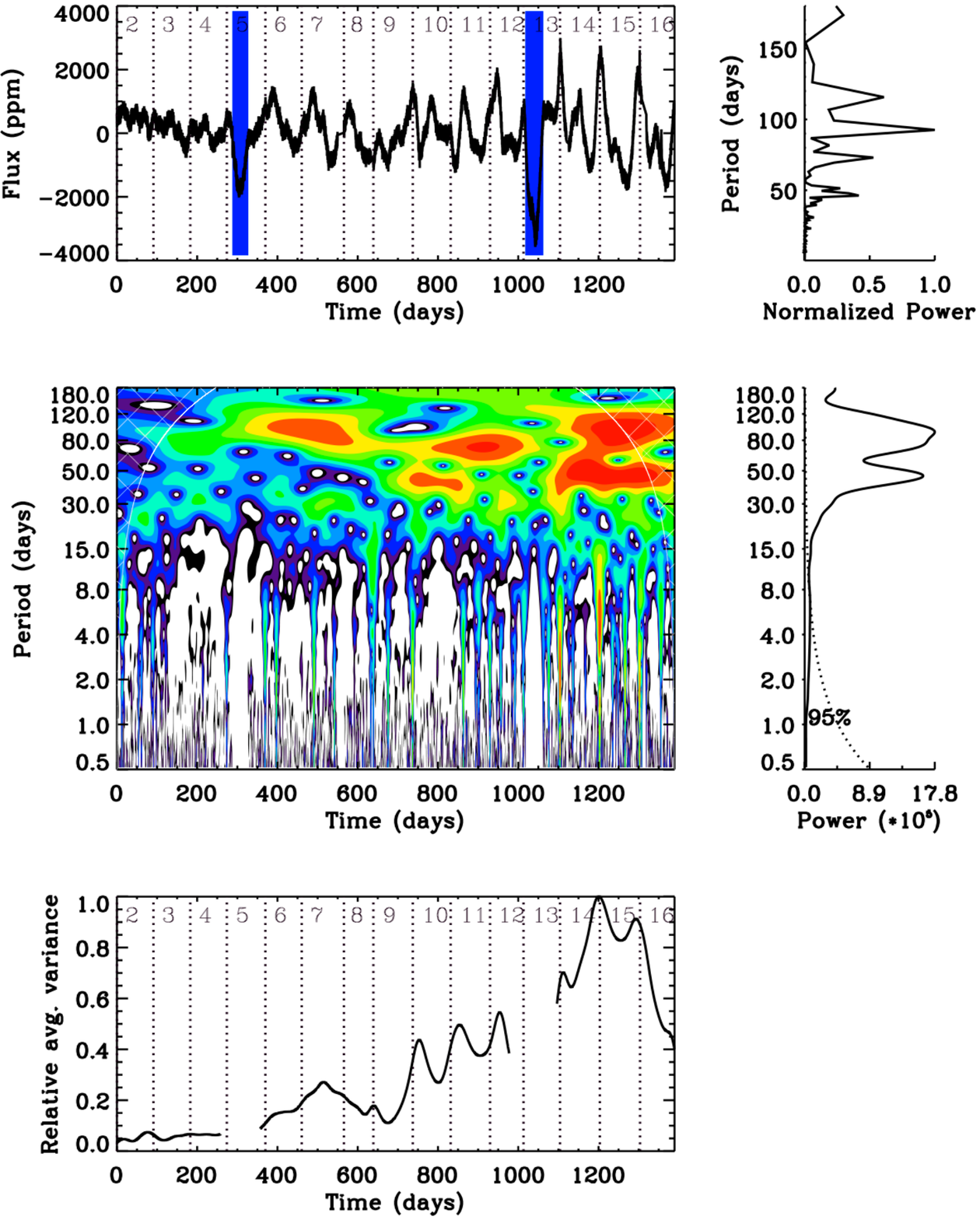}
\caption{Top left panel: Long-cadence light curve of KIC~8561221 starting on June 20, 2009 where dotted lines indicate the transitions between the observing quarters.  Top right panel associated power density spectrum expressed in period between 0.5 and 180 days. The blue shaded regions were removed in the analysis. Middle left panel: Wavelet power spectrum computed using a Morlet wavelet between 0.5 and 180 days on a logarithmic scale. Middle right panel:  Global power spectrum as a function of the period of the wavelet. The dotted line is the confidence level corresponding to a $95\%$ probability. Bottom panel: The scale-averaged time series reconstructed from the periods in the range 15 to 150 days. The discontinuities at Q5 and Q13 are a consequence of the rejected regions (enlarged to avoid border effects).}
\label{Rotation}
\end{figure*}

We have used Simple Aperture Photometry (SAP) time series \citep[e.g.][]{ThompsonRel21} that were corrected for different instrumental perturbations --outliers, jumps, and drifts-- using the procedures described in \citet{2011MNRAS.414L...6G} and high-pass filtered by a 2-day triangular smoothing function. 

To complement the seismic studies, we measured the surface rotation rate from the modulations in the light curve following previous analyses on space-borne photometry \citep[e.g.][]{2009A&A...506...41G,2011A&A...534A...6C,2011ApJ...733...95M}. To do so,  we have used SAP LC photometry for almost the full {\it Kepler} mission from Q2 to Q16, i.e. from June 20, 2009 till April 08, 2013 corrected following  \citet{2011MNRAS.414L...6G}. The data corresponding to Q0 and Q1 were rejected because of instrumental instabilities. We have  filtered the data using a double triangular smoothing function with a 200 days width. The LC flux is shown in the top panel of Fig.~\ref{Rotation}. The blue shaded regions during Q5 and Q13 mark two regions that were rejected in the analysis because they seemed to be affected by instrumental instabilities. Unfortunately, in the case of KIC~8561221, the data corrected using Pre Data Conditioning multi-scale Maximum a Posteriori methods \citep[PDC-msMAP, e.g.][]{ThompsonRel21} are not 
reliable because they are high-pass filtered with a limit at around 21 days.

\subsection{Surface rotation period and magnetic activity}
To determine the average rotation rate of the stellar surface we look for any modulation in the light curve induced by the dimming produced by spots crossing the visible stellar disk. As can be seen in the top left panel of Fig.~\ref{Rotation}, there is a regular long-period modulation clearly visible in the light curve. This modulation seems to be due to variations in the brightness of KIC~8561221 because the contamination of the field is very small, 0.000264, and visual inspection of the UKIRT (\url{http://keplerscience.arc.nasa.gov/ToolsUKIRT.shtml}) high-resolution images does not show any other star that could pollute the photometric signal. Finally, the modulation can be seen at the same level for all pixels within the photometric mask. 


To study the rotation period we use two complementary techniques which have been described in \citet{2013arXiv1307.4163G}{: the study of the Global Power Spectrum \citep[e.g.][]{2010A&A...518A..53M} and the autocorrelation of the temporal signal \citep[e.g.][]{2013MNRAS.432.1203M}. Unfortunately, as said in the previous section, we were not able to use PDC-msMAP corrected data and we have only used the data corrected following \citet{2011MNRAS.414L...6G}. 

The Global Power Spectrum (GPS) represented in the middle right-hand panel of Fig.~\ref{Rotation} is a periodogram computed by projecting into the time domain the wavelet power spectrum shown in the middle left-hand panel. We use the Morlet wavelet, i.e., a Gaussian envelope with a varying width \citep{1998BAMS...79...61T}, which we prefer to the analysis of the low-frequency end of the standard periodogram (shown in the top right-hand panel in Fig.~\ref{Rotation}) because it allows us to verify if the dominant features at low frequency are the consequence of isolated events (related to instrumental perturbations) or if they are present during all the observations (see middle left-hand panel of Fig.~\ref{Rotation}). Although there is an increase of the photometric modulation after Q9, a periodic signal is visible in the whole light curve. Fitting three Gaussian functions to the global power spectrum provides a rotation period, $P_{\rm rot}=91.1 \pm 12.5$ days, with a secondary period at 46.0 $\pm$ 3 days, and a third at 73.6 $\pm$ 2.5 days (here we quote the half width at half maximum, of the fitted functions as the uncertainty). The period of 46 days corresponds to the second harmonic of the rotation period. Indeed the time frequency diagram shows that the $\sim$91-day period is seen at several moments during the observations (Q6 and Q7, Q10 to Q12, and during Q14 to Q16). However, there is considerable dispersion in the exact period which is reflected in the large uncertainty we obtain. This could be the signature of surface differential rotation.

The analysis of the autocorrelation function of the light curve --shown in Fig.~\ref{full_auto}-- confirms the value of the rotation period obtained by the GPS. The maximum peak of the autocorrelation is at 90.9 $\pm$ 11.7 days (uncertainty computed as the half width at half maximum of the peak).
\begin{figure}[!htb]
\centering
\includegraphics[scale=0.36]{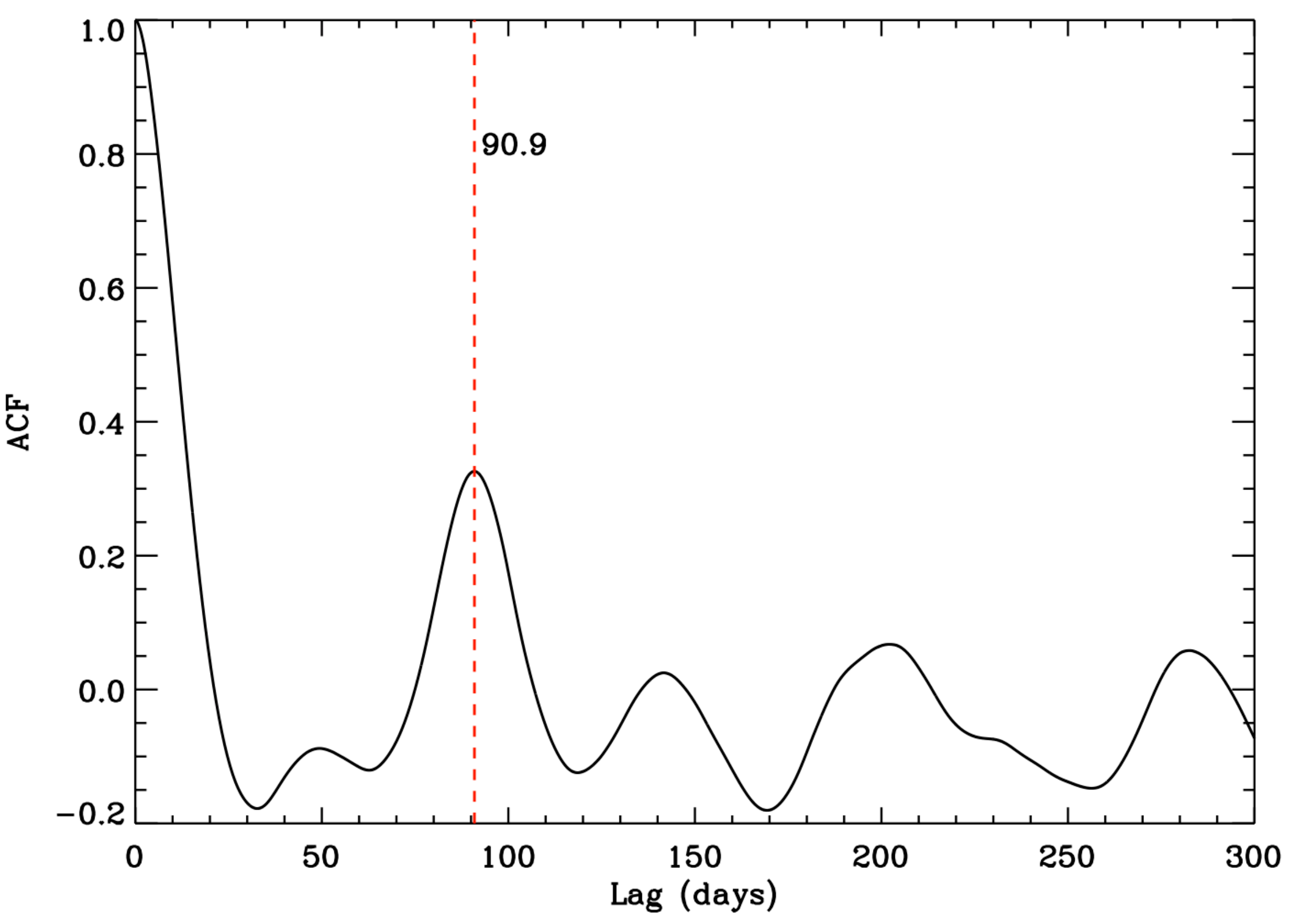}
\caption{Autocorrelation function (ACF) of the light curve showed in the top left-hand panel of Fig.~\ref{Rotation}. The maximum of the ACF caused by rotation is marked by a vertical red dashed line (90.9 $\pm$ 11.7 days).}
\label{full_auto}
\end{figure}


The observed $P_{\rm rot}$ is suspiciously close to the length of a {\it Kepler} quarter. To decide whether or not this measurement is of stellar origin, we need to check if it is in phase with the quarters. If we take a closer look at the light curve, for example between Q6 and Q8 (see top left-hand panel in Fig.~\ref{Rotation} and Fig.~\ref{auto_zoom}), the 
 maxima of the modulation are not aligned with the quarters and the ends of each quarter are well stitched together without any significant jumps between them (see Fig.~\ref{auto_zoom}). The ACF of just these quarters provides $P_{\rm rot} \sim$~93 days although the average length of the quarters is $\sim$ 82 days. Similar results are obtained when other sections of the light curve are analyzed separately. We therefore conclude that the retrieved $P_{\rm rot}$ is probably of stellar origin.  

To theoretically estimate the surface rotation we adopted a 1.5 solar mass reference model at [Fe/H]=0 and evolved it to $\log g$ = 3.62 using the input physics and angular momentum evolution model described in \citet{2013ApJ...776...67V}.  We assumed rigid rotation and a range of initial conditions chosen to bracket the observed distribution of rotation rates in intermediate aged open cluster stars of this mass range.  Angular momentum loss was included but does not have a significant effect on the main sequence; however, these models do experience significant post-main sequence angular momentum loss that was included in the calculations.  With these models we predict a surface rotation rate in the range 35 to 43 days. This range reflects the range of main sequence rotation rates for normal stars and assumes rigid rotation at all times.  If the progenitor had a lower mass the expected rotation periods would increase. For example, the surface rotation increases to the range  58 to 63 days by decreasing the mass to 
1.4 $M_\odot$.  Core-envelope decoupling in the post-main sequence would decrease the predicted surface rotation, although \citet{2013ApJ...776...67V} found that this had only a modest impact on model predictions; if the main sequence precursors had rapidly rotating embedded cores the surface rotation would be above the predicted range.

A lower rotation rate could also be obtained if the precursor was a chemically peculiar star or if we consider any extra magnetic braking at the surface of the star. As we will see later, KIC~8561221 seems to still have a magnetic field in action at the surface with a possible observational signature of a dynamo at this stage of evolution. 

\begin{figure}[!htb]
\centering
\includegraphics[scale=0.36]{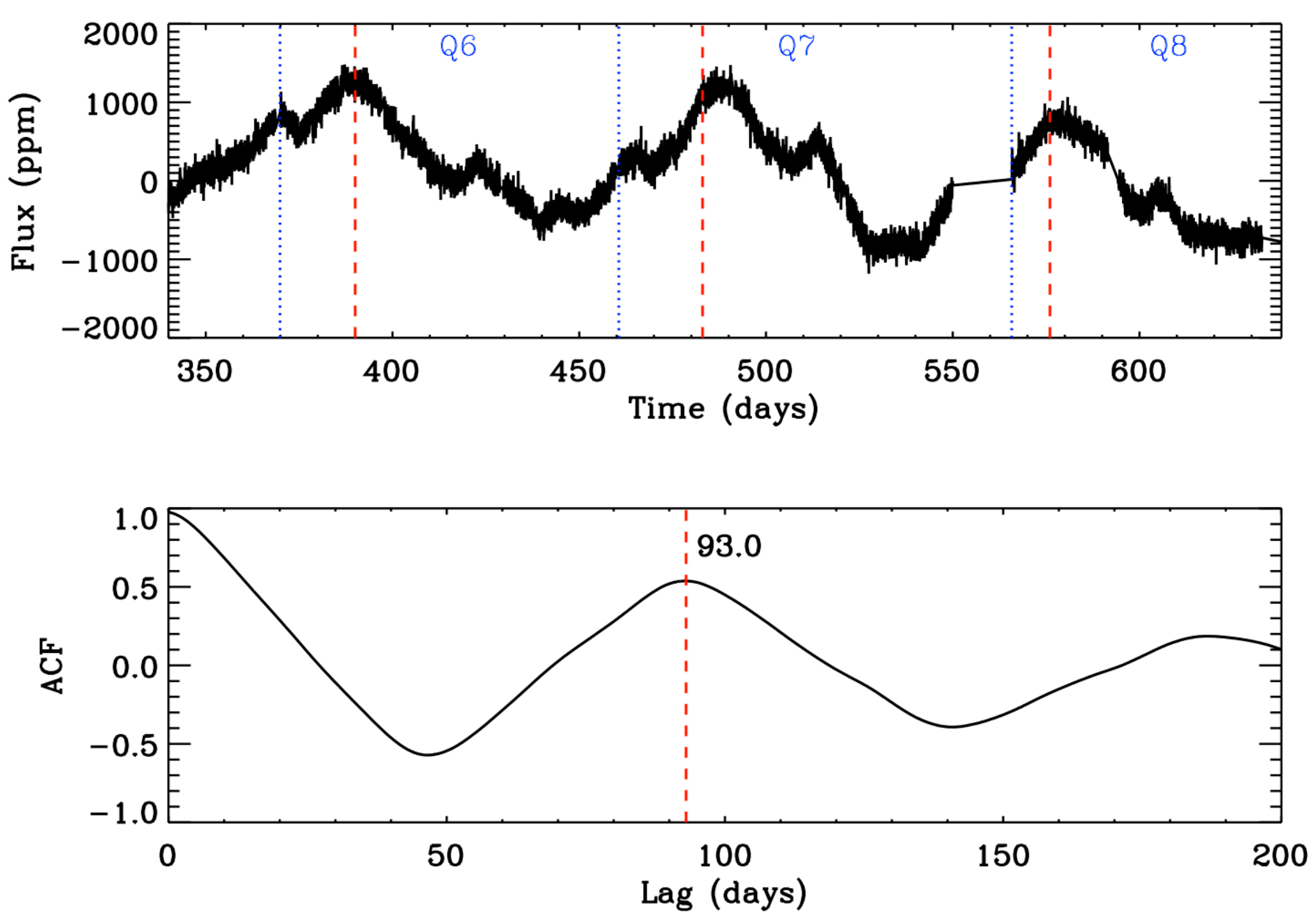}
\caption{Zoom of the light curve of KIC~8561221 (top) and ACF of this part of the light curve. The position of maximum of the ACF, 93.0 days, is marked by a vertical red dashed line in the bottom panel. The red-dashed lines in the light curve (top panel) are the corresponding period intervals, which match the repeating flux modulations. The ends of each quarter are indicated by the blue dotted lines. }
\label{auto_zoom}
\end{figure}


The presence of spots on the surface of the star demonstrated by the low-frequency modulation of the photometric signal shows that the star is active. We can define a photometric activity index, $S_{\rm ph}$ directly from the light curve in order to quantify the activity and to be able to compare it with other {\it Kepler} targets and the Sun (e.g. Mathur et al. submitted). To do so, we divide the light curve into subseries of $5 \times P_{\rm rot}$ (to have enough realizations of the rotation per subseries) and compute the standard deviation of each segment. We have verified that the results do not change when we use subseries of $3 \times P_{\rm rot}$. The activity index, $\langle S_{\rm ph} \rangle$, will be the average of the individual standard deviations of each subseries. We prefer to use the average instead of the median because, if there is an on-going magnetic activity cycle, the median underestimates the final results and because we only have three independent realizations of $5 \times P_{\rm rot}$. Finally, we subtract the photon noise (19.9 ppm) computed following the procedure described in \citet{2010ApJ...713L.120J}. For KIC~8561221 we obtain  $\langle S_{\rm ph} \rangle =637.0$ $\pm$ 5.1 ppm with a minimum activity of 473 $\pm$ 3.5 ppm and a maximum of 965 $\pm$ 7.2 ppm.  As a comparison, we can compute the same activity index for the Sun, $\langle S_{\rm ph,\odot} \rangle$ using the sum of the red and green channels of the Sun SpectroPhotoMeters (SPM) of the VIRGO instrument \citep{1995SoPh..162..101F} aboard the Solar and Heliospheric Observatory (SoHO) \citep{DomFle1995}. Indeed --as recently demonstrated by \citet{2013ApJ...769...37B}-- the use of these two colors of  VIRGO/SPM constitutes a good approximation to the {\it Kepler} photometric bandpass. In the solar case --using 16 years of quasi-continuous measurements-- we found that $\langle S_{\rm ph,\odot} \rangle =166.1$ $\pm$ 2.6~ppm with a variation between the minimum and the maximum of 89.0 $\pm$ 1.5 and 258.5 $\pm$ 3.5 ppm respectively. Therefore the changes in the photospheric activity of KIC~8561221 are around 3.5 times higher than the one exhibited by the Sun, although it is a more evolved star (see Sect.~\ref{Sec_Model}). 

To uncover the possible existence of a magnetic activity cycle in KIC~8561221, we compute the scale-averaged time series, i.e., the projection into the time domain of the wavelet power spectrum around the  time scales of the rotation period  \citep[an application of this methodology to the Sun can be seen in][]{2013JPhCS.440a2020G}. This is done to ensure that the temporal variations we are computing are only related to the stellar spots, and not to any other instrumental or stellar feature at different time scales (see bottom panel in Fig.~\ref{Rotation}).  However, we have followed a conservative approach by using a large range of periods (between 15 and 150 days). It is also important to notice that we have removed additional data in the vicinity of the rejected observations at Q5 and Q13 because we need to take into account border effects on the reconstruction of the signal.

 There is a clear increase in the magnetic activity at the end of Q9 with a maximum around Q14 and Q15, followed by a slow decline.
  We can therefore argue that despite its late stage of evolution (see Sect.~\ref{Sec_Model}) it is very likely that a magnetic activity cycle (probably a dynamo) is in action in KIC~8561221.

\subsection{Background parameters}
The convective background of KIC~8561221 has been studied by fitting a simple 3-component model, $B(\nu)$ to the power spectrum. The first component is the photon noise, $W$, which dominates the frequency-independent noise at high frequency. The second component is a Harvey-like profile \citep{harvey85} to model the granulation contribution:

\begin{equation}
B_{\rm g}(\nu)=\frac{4 \tau_{\rm g} \sigma^2_{\rm g}}{1+(2 \pi \tau_{\rm g} \nu)^{\alpha_{\rm g}} } \;,
\label{eq_back}
\end{equation}
in which $\tau_{\rm g}$ is the characteristic time scale of the granulation, $\sigma_{\rm g}$ is its amplitude, and $\alpha_{\rm g}$ is an exponent characterising its temporal coherence. 

The third background component models the low-frequency part of the power spectrum dominated by activity. For this we use a power law $B_a(\nu)=P_a \nu^{e_a}$ .

To obtain the best fit we have followed the same procedure as in \citet[][]{2011A&A...530A..97B}. The results of the fit are  $\tau_g (s)$ = 1112 $\pm$ 21,  $\sigma_g$ = 102.2 $\pm$ 0.5 ppm, and  $\alpha_g$ = 2.41 $\pm$ 0.02.

In addition, we have computed the effective granulation timescale as defined in \citet{2011ApJ...741..119M}. We obtained a value of $\tau_{\rm eff}=1415 \pm 19$ s, which is in perfect agreement with the expected value from the relation inferred for red-giant stars by \citet{2011ApJ...741..119M}. Therefore, it seems that there are no differences in the convection timescales between this star and those having dipole modes with normal amplitudes.


\begin{figure}[!htb]
\centering
\includegraphics[scale=0.5]{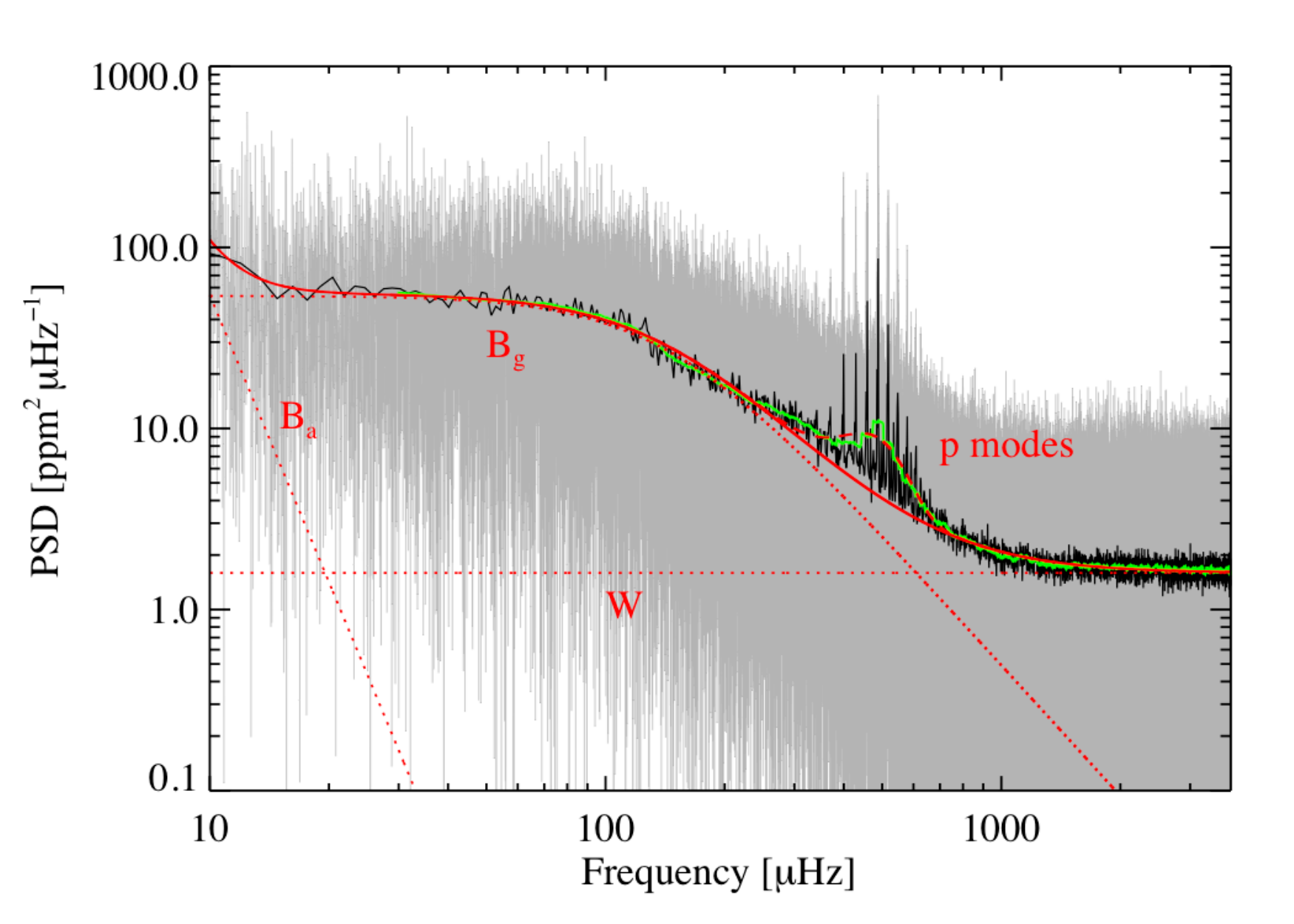}
\caption{Power density spectrum of KIC~8561221 Short-cadence data at full resolution (grey curve) and after rebinned by a factor of 100 (black curve). The green line shows the spectrum smoothed by a box car with a width equal to the mean large separation $\Delta \nu$. The fitted background is the solid red line, and its three components (W, Bg, Ba) are plotted as red dotted lines. P-mode power excess, fitted as a Gaussian profile, is also represented with a red dashed line.}
\label{Background}
\end{figure}





\subsection{Seismic characterization}
\label{fitting}

We determined the global parameters of the acoustic modes of the star using the A2Z pipeline \citep{2010A&A...511A..46M}, one of the extensively tested tools to reliably extract those parameters \citep[see e.g.][]{2011A&A...525A.131H}.
The mean large separation was obtained by computing the power spectrum of the power spectrum. This led to $\langle \Delta \nu \rangle$ = 29.88\,$\pm$\,0.80 $\mu$Hz. We fitted the p-mode bump with a Gaussian function to measure the frequency at maximum power and the maximum amplitude: $\nu_{\rm max}$ = 490\,$\pm$\,24 $\mu$Hz and $A_{\rm max}$ = 6.8\,$\pm$\,0.24\,ppm.

A Bayesian approach using a Markov Chain Monte Carlo algorithm  \citep{Benomar2009a} was used to globally extract the characteristics of the individual p modes in KIC~8561221. The priors  are similar to the ones defined by \citet{2013ApJ...767..158B}. However, we  directly fit each dipole mixed mode instead of using the coupled oscillator model. The results were compared and validated using standard Maximum Likelihood Estimator (MLE) fits.

\begin{figure*}[!htb]
\centering
\begin{tabular}{cc}
\includegraphics[scale=0.40]{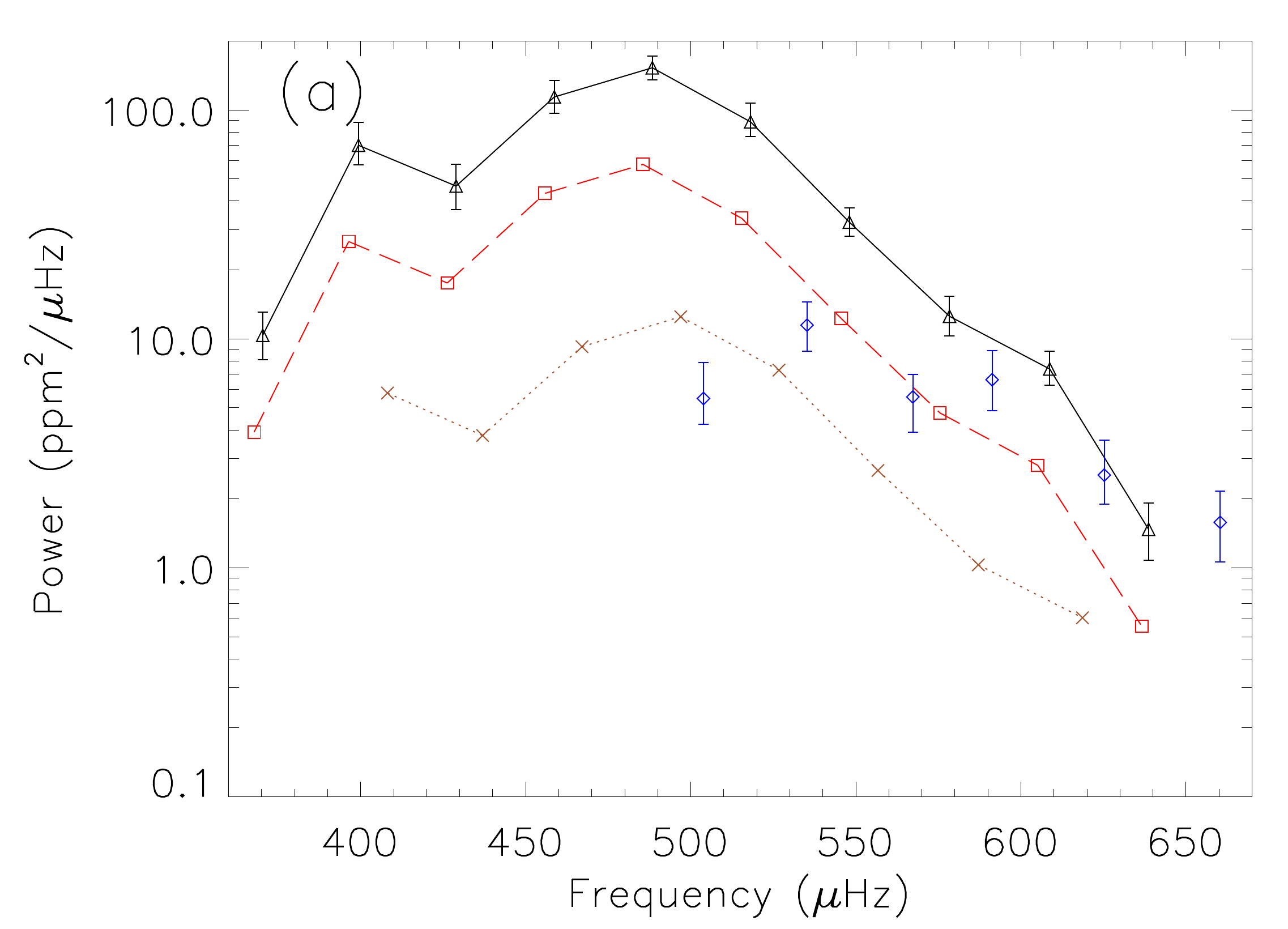}&
\includegraphics[scale=0.40]{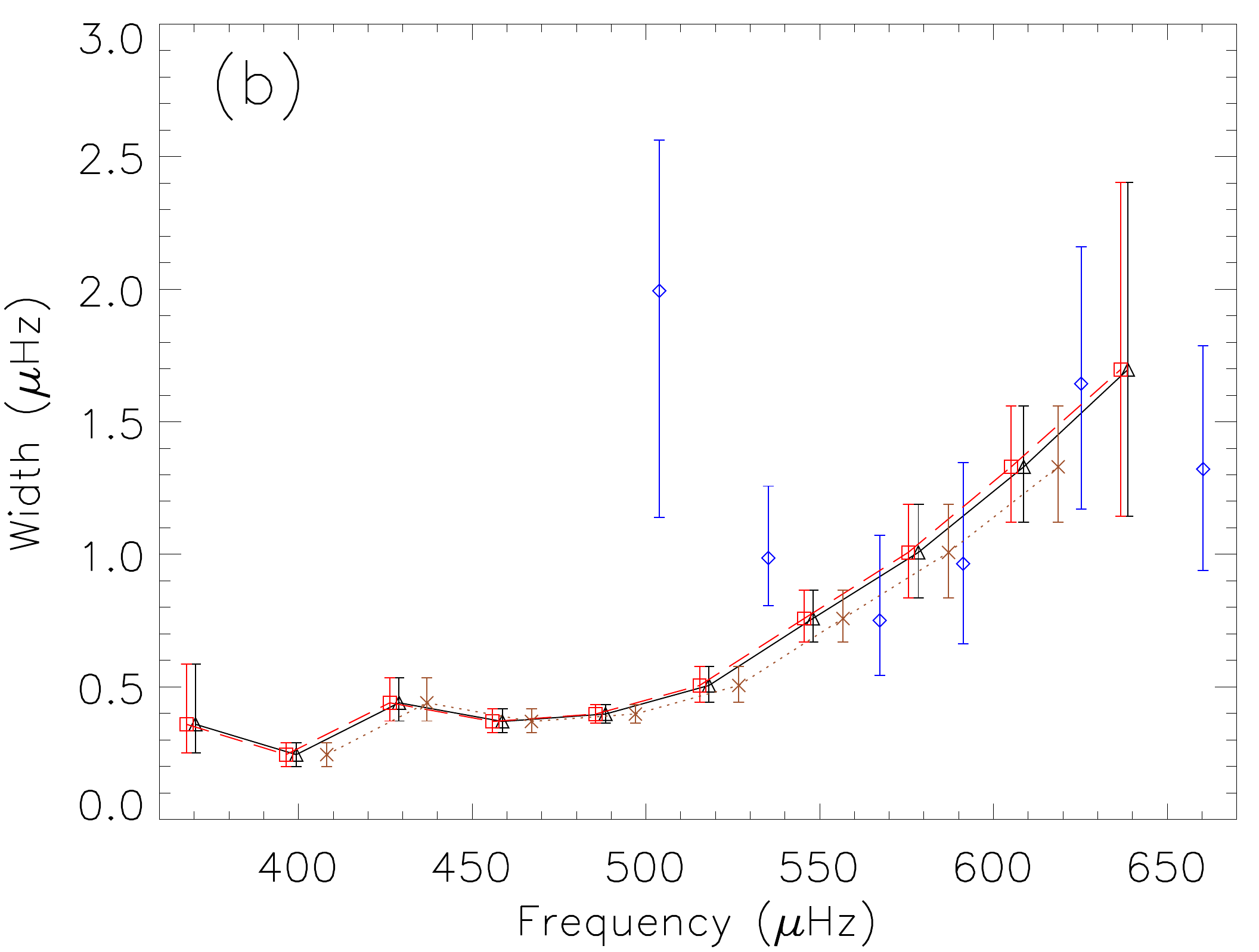}\\
\includegraphics[scale=0.40]{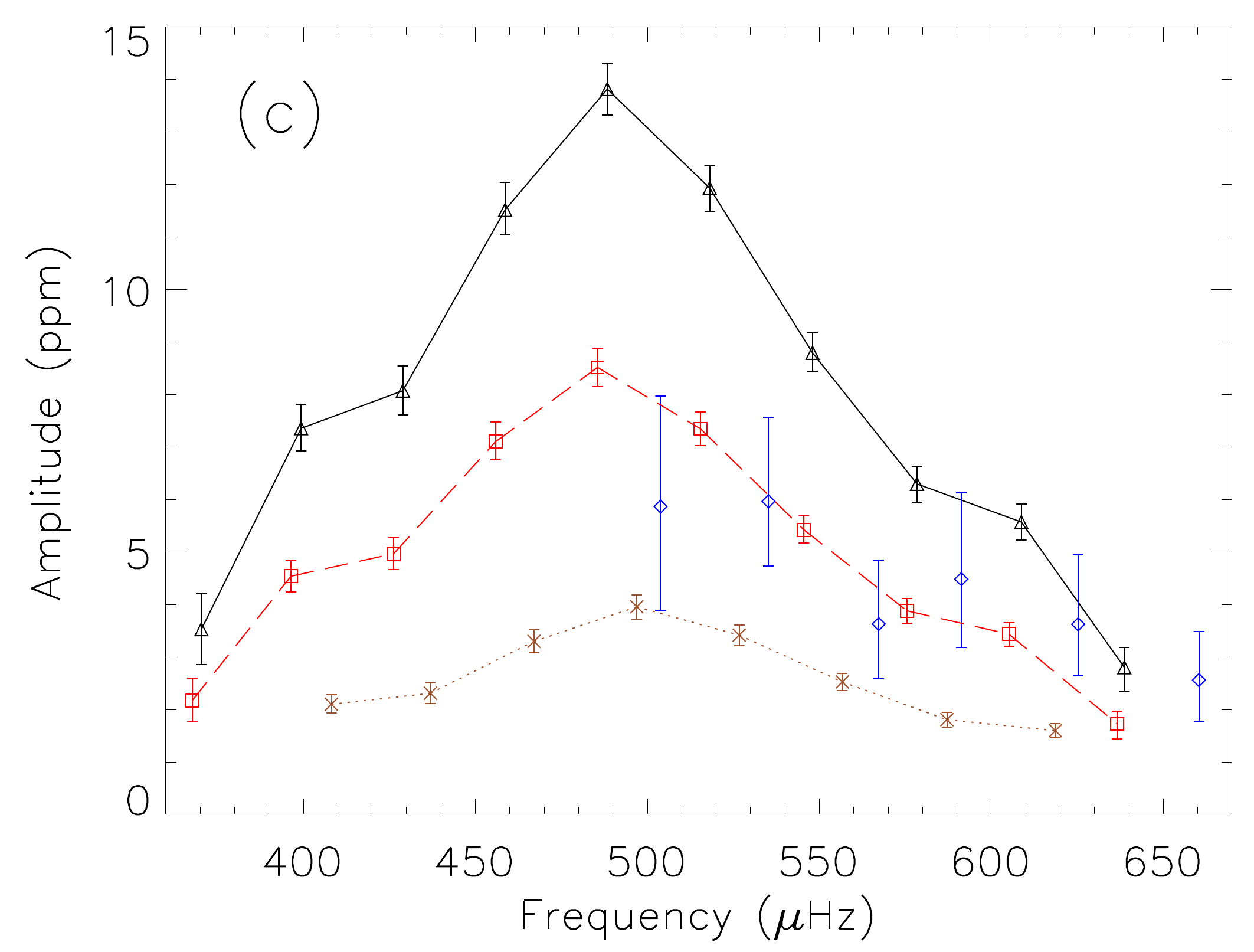}&
\includegraphics[scale=0.40]{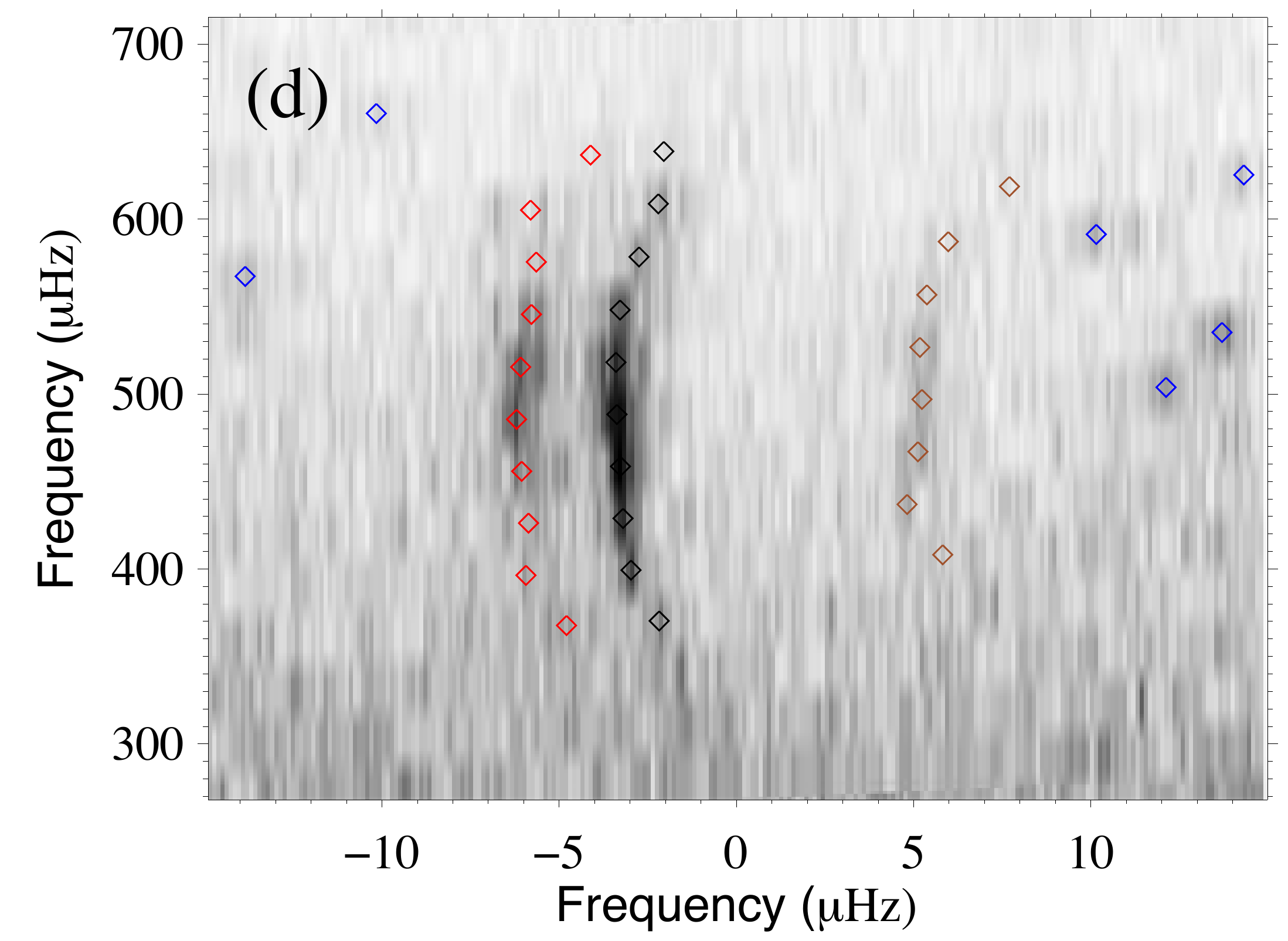}
\end{tabular}
\caption{\label{fig_HLA} Height, i.e. power (a), linewidth (b), and amplitude (c) of the fitted modes following  ``case 3''. See the text for further details. Colors correspond to different degrees: $\ell=0$ (black triangles), $\ell=1$ (blue diamonds), $\ell=2$ (red squares), and $\ell=3$ (brown crosses). To indicate that all parameters of the dipole modes were fitted independently, while the amplitude ratios and linewidths of the $\ell$=2 and $\ell$=3 modes were fitted together, the points for the $\ell$=0, 2 and 3 modes are connected by  black continuous, red dashed, and brown dotted lines respectively. Panel (d) shows the \'echelle diagram with the measured frequencies (diamonds) superimposed on the power spectrum (gray scale).
}
\end{figure*}

Modes of degree $\ell=0, 1,2$ and $3$ are identified in the power density spectrum. Modes of degree $\ell=1$ clearly behave as mixed modes although their amplitudes are rather low. Each mode was fitted using a sum of symmetric Lorentzian profiles, one per $m$-component of the multiplet. We assumed that the $\ell=2$ and $\ell=3$ modes are p-dominated modes. Indeed  $\ell \geq 2$ pressure and gravity modes are weakly coupled due to the relatively wide evanescent region between the p- and g-mode cavities \citep[e.g., Eq. 16.51 of ][]{1989nos..book.....U}.  This also means that the height of $\ell \geq 2$ roughly scales with the height of $\ell=0$, by a factor $V_l$ (considered as a free parameter), which is the mode visibility.


The rotational splitting, $\delta\nu_s(\ell)$, was included under the assumption that the inclination angle is the same throughout  the star (the interior spins with the same inclination angle as that of the surface). However, we have tried different hypotheses on the splittings, $\delta\nu_s(\ell)$, and on the mode widths, $\Gamma(\ell)$, in order to evaluate the robustness of the results. A total of five scenarios were explored and their characteristics are summarised in Table~\ref{tab:cases}.  The one with the highest degree of freedom (Case 1) contains the following assumptions: 
	\begin{itemize}
		\item[$\bullet$] $\delta\nu_s(\ell=1)$ is different for each $\ell=1$ mixed mode.
		\item[$\bullet$] $\delta\nu_s(\ell=2)$ is the same for all $\ell=2$ modes.
		\item[$\bullet$] $\delta\nu_s(\ell=3)$ is the same for all $\ell=3$ modes.
		\item[$\bullet$] $\Gamma_n(\ell=0)$ is a smooth function of the radial order.
		\item[$\bullet$] $\Gamma_n(\ell=2)$ and $\Gamma_n(\ell=3)$ are interpolated at the mode frequencies, using the $\Gamma_n(\ell=0)$ curve.
		\item[$\bullet$] $\Gamma(\ell=1)$ and  heights $H(\ell=1)$ are different for each mixed mode. 
	\end{itemize}
Case 2 uses the same assumptions as case 1 but $\delta\nu_s(\ell=2)=\delta\nu_s(\ell=3)$, while case 3 assumes in addition that $\Gamma(\ell=1) \leq \Gamma(\ell=0)$, as is expected for mixed modes. We did not, however, use a strict boundary. The prior is uniform when $\Gamma(\ell=1) \leq \Gamma(\ell=0)$ and Gaussian otherwise, with a standard deviation of $0.2\times\Gamma(\ell=0)$. Another scenario (case 4) uses the same assumptions as case 2 but includes only the four highest signal-to-noise $\ell=2$ modes (over a total of 11 detected $\ell=2$ modes). Finally, the last scenario (case 5) is similar to case 4 but adds the condition $\Gamma(\ell=1) \leq \Gamma(\ell=0)$. Case 4 and case 5 evaluate the sensitivity to the model of the $\ell=2$ and $\ell=3$ rotational splittings.

\begin{table*}[!htb]
\begin{center}
\caption{Fitting scenarios studied to check the robustness of the result. When the value is not changed compared to the preceding case it has been indicated by ``=''.}
\begin{tabular}{ccccccc}
\hline
\hline
Case & $\delta\nu_s(\ell=1)$ & $N_{fitted}(\ell=2)$ &$\delta\nu_s(\ell=2, 3)$ & $\Gamma(\ell=0)$ & $\Gamma(\ell=1)$ & $\Gamma(\ell=2, 3)$ \\
\hline
1 & independent & All &$\delta\nu_s(\ell=2) \neq \delta\nu_s(\ell=3)$ & independent & independent &interpolated from $\Gamma(\ell=0)$ \\
2 & = & = & $\delta\nu_s(\ell=2) = \delta\nu_s(\ell=3)$& = & = & = \\
3 & = & = & = & = &  $\Gamma(\ell=1) \leq \Gamma(\ell=0)$ & =\\
4 & = & 4 highest SNR & = & = & independent & = \\
5 & = & = & = & = & $\Gamma(\ell=1) \leq \Gamma(\ell=0)$ & =\\
\hline
\label{tab:cases}
\end{tabular}
\end{center}
\end{table*}

\begin{figure*}[!htbp]
\centering
\includegraphics[scale=0.70]{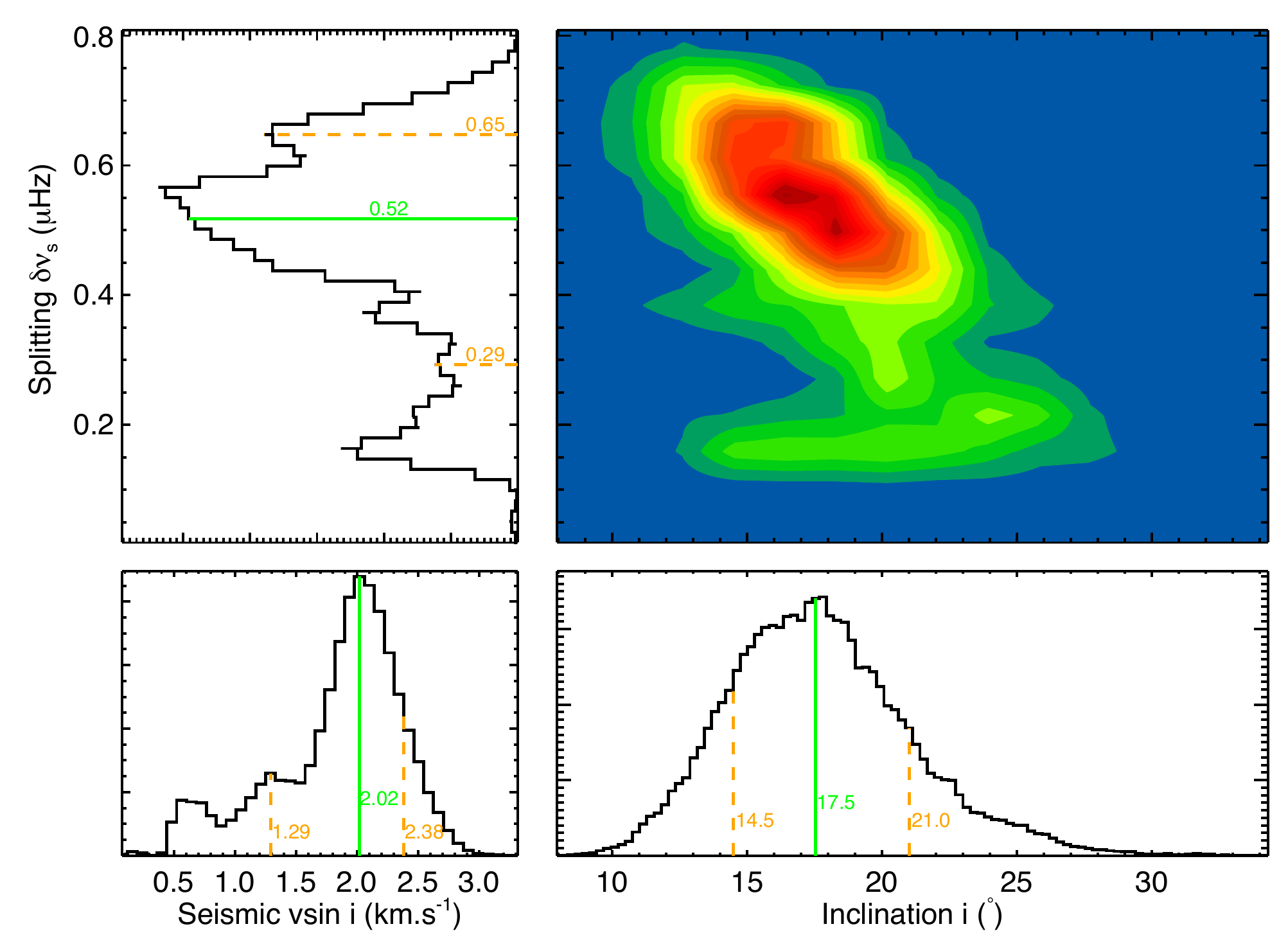}
\caption{Correlation map between $\ell > 1$ splitting and the inclination angle (upper right) for case 3, as well as their PDF (upper left, lower right respectively). The seismic $v \sin i$ (lower left) is derived using the radius and effective temperature of the RADIUS pipeline, and the p-mode splitting showed in the top-right panel. The solid green lines correspond to the median of the distributions and the orange dash lines indicate the 68$\%$ confidence interval.}
\label{dist_rot_ang}
\end{figure*}

All the scenarios give frequencies that are consistent within $1\sigma$ for all fitted modes, which reinforces the robustness of the determination of the frequencies within this set of sensible choices for the fitting approach. However, we notice that in the cases 4 and 5 the uncertainties on the estimated frequencies of the $\ell=0$ modes are larger when the neighbouring $\ell=2$ is not fitted. This is expected because these modes are partially blended.  
Fig.~\ref{fig_HLA}a shows the height (power density) of the modes as a function of frequency for the reference fit: case 3. 

Usually, for {\it Kepler}, the visibility of $\ell=1$ modes, $V_{\ell=1}$, is $\simeq1.5$ times that of $V_{\ell=0}$, while $V_{\ell=2} \simeq 0.5$, and $V_{\ell=3} \simeq 0.08$ \citep[e.g.][]{2011A&A...531A.124B}. Here, it is clear that $V_{\ell=1}$ is much lower than expected. Moreover, while in other stars the relative height remains constant over all the observed frequency range,  this is not the case for KIC~8561221. In this star the dipole modes around $\nu_{\rm{max}}$ are as weak as $\ell=3$ modes but the $V_{\ell=1}/V_{\ell=0}$ ratio increases to about one towards the high-frequency modes. Dipole modes at frequencies lower than $\nu_{\rm{max}}$ are too weak to be measured. Observed dipole mixed modes are close to the $\ell=1$ p-mode home ridge (see the \'echelle diagram shown in Fig.~\ref{fig_HLA}d). Therefore, only modes with the lowest inertia (meaning efficiently trapped in the p cavity) are visible.
Finally we notice that $V_{\ell=2}$ is lower than expected ($0.38 \pm 0.03$, see top panel in Fig.~\ref{fig_visibilityl2}). The difference could be due to the non-accounted visibility of the g-dominated quadrupole modes that we were not able to measure \citep[see for more details][]{2012A&A...537A..30M}. $V_{\ell=3}$ is normal ($0.08 \pm 0.01$, see bottom panel in Fig.~\ref{fig_visibilityl2}). 

Fig.~\ref{fig_HLA}b shows the mode width as a function of frequency. High-frequency dipole modes have widths similar to modes of different degree but similar frequency. However, at lower frequency, widths apparently increase, which must be significant because the prior in the case-3 fit (shown in Fig.~\ref{fig_HLA}b) penalizes the width of the dipole modes to be significantly greater than for the radial modes.  A closer analysis of the power spectrum indicates that modes around $504$ $\mu$Hz and $535$ $\mu$Hz are indeed broad and correctly fitted. The amplitude of the dipole modes, determined by the relation $A = \sqrt{\pi \rm{H} \Gamma}$, remains in overall lower than the $\ell=0$ amplitudes (Fig.~\ref{fig_HLA}c).

The different scenarios all yield inferred inclination angles that are consistent within $1\sigma$, spreading from $13^\circ$ to $29^\circ$. This suggests that the star is seen from a low angle close to the pole (see Appendix~\ref{Annex} for a detailed discussion). However, with such a low inclination angle, measurements of the rotational splitting become less precise, especially for the dipole modes \citep[e.g.][]{2003ApJ...589.1009G,BalGar2006}. This was verified by looking at the probability distributions of the $\delta\nu_s(\ell=1)$ that are highly multimodal. The splittings of the $\ell$=1 modes have median values of several $\mu$Hz and have a standard deviation of the same order (a computation of the average splitting does not improve the estimates). The $\ell=2-$ and $\ell=3-$mode rotational splittings are much better constrained, although they depend on the hypothesis we included in the fitting model. Thus,  in the first case, $\delta\nu_s(\ell=2)=0.69 \pm 0.10$ $\mu$Hz and $\delta\nu_s(\ell=3)=0.16 \pm 0.08$ $\mu$Hz are inconsistent and non gaussian. Indeed, such a solution would have implied that the kernel of $\ell=2$ and $\ell=3$ modes are very different as well as a very steep radial differential rotation}. However, when we assume $\delta\nu_s(\ell=2)=\delta\nu_s(\ell=3)$ (cases 2 to 5), the rotational splitting becomes consistent within $1\sigma$.  The  median values we obtain are ranging between 0.3 and 0.6 $\mu$Hz. The lower limit is close to twice the inferred surface rotation rate of about 91 days and it would imply an increase of the internal rotation rate as has already been observed in other evolved stars \citep{2012Natur.481...55B,2012ApJ...756...19D}.

Fig.~\ref{dist_rot_ang} shows the correlation maps between the rotational splitting of $\ell > 1$ modes and the inclination axis (case-3 fit). The splitting shows a multi modal distribution and, in contrast, the maximum probability for the inclination axis is very well defined.  This figure also shows the probability density function (PDF) of the seismic $v \sin i$, calculated using the rotational splitting, and the effective temperature and the radius obtained by the RADIUS pipeline (assuming gaussian uncertainties, see Table~\ref{table5} and Sect.~\ref{Sec_Model}). 


\begin{figure}[!htb]
\centering
\includegraphics[scale=0.37,trim=8cm 2cm 6cm 2cm]{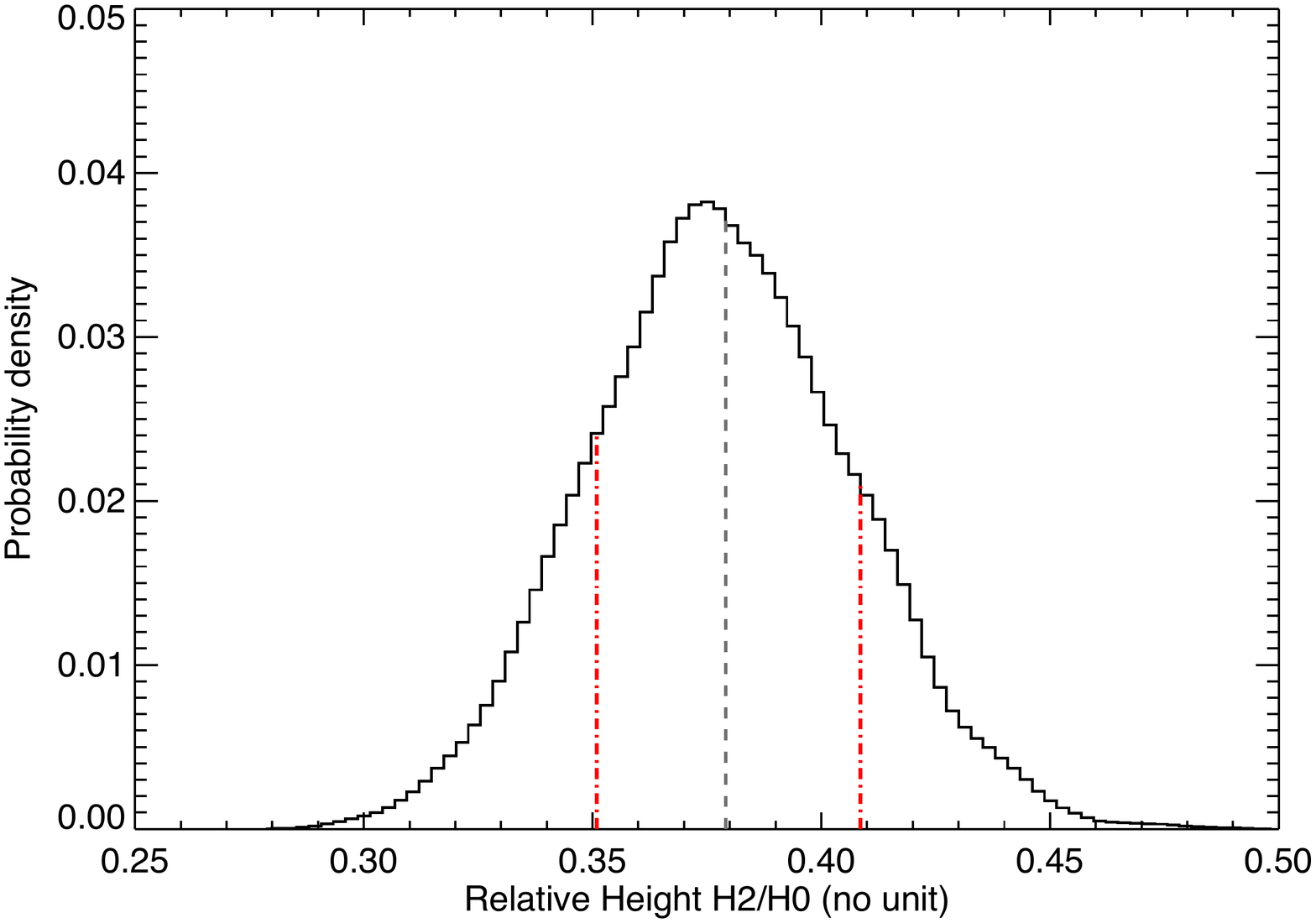}
\includegraphics[scale=0.37,trim=6cm 2cm 4cm 2cm]{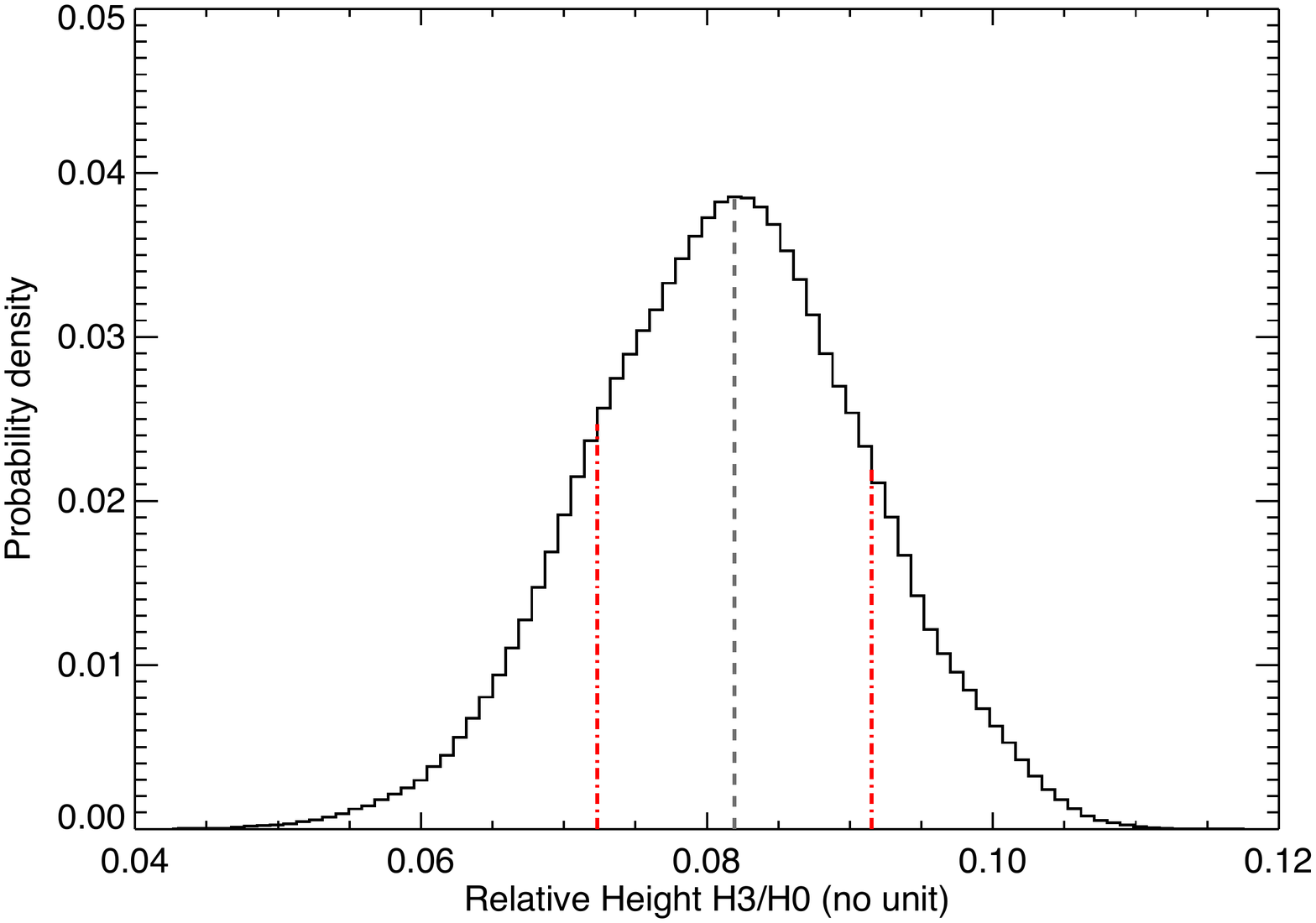}
\caption{Probability Density Function (PDF) of the $\ell$=2  (top) and $\ell$=3 (bottom) mode visibility. The dashed line corresponds to the median of the distribution and the red dot-dashed lines indicate the 68$\%$ confidence interval around it.}
\label{fig_visibilityl2}
\end{figure}


Concerning the extracted heights and widths, the values obtained for  $\ell$ = 0, 2 and 3 modes are very stable and they are consistent among all the different scenarios. However, due to the low signal-to-noise ratio of the dipole modes, their heights and widths are barely measurable, and the uncertainties increase dramatically towards low frequencies.


\begin{table*}
\begin{center}
\caption{Frequencies, amplitudes, and linewidths for all the modes computed with the MCMC for case 3. }
\begin{tabular}{ccccccccc}
\hline
\hline
$l$ & Frequency ($\mu$Hz) & $\sigma$ & Linewidth ($\mu$Hz) & $+\sigma$ & $-\sigma$ & Height (ppm)& $+\sigma$ & $-\sigma$ \\
\hline
 0    &   370.33 & 0.06  &     0.36 & 0.11 & 0.23  &    10.37 &  2.26 & 
 2.73\\
0   &   399.34 & 0.02  &     0.24 & 0.04 & 0.05  &    70.03 & 12.57 & 
18.21\\
0    &   428.91 & 0.02  &     0.44 & 0.07 & 0.09  &    46.48 &  9.78 & 
11.39\\
0    &   458.64 & 0.01  &     0.37 & 0.04 & 0.05  &   113.92 & 17.26 & 
20.66\\
0    &   488.34 & 0.00  &     0.40 & 0.03 & 0.03  &   152.80 & 17.46 & 
18.89\\
0   &   518.12 & 0.03  &     0.51 & 0.06 & 0.07  &    88.67 & 12.32 & 
18.29\\
0    &   548.03 & 0.05  &     0.76 & 0.09 & 0.11  &    32.36 &  4.34 & 
 5.07\\
0    &   578.37 & 0.09  &     1.01 & 0.17 & 0.18  &    12.55 &  2.25 & 
 2.78\\
0    &   608.72 & 0.11  &     1.33 & 0.21 & 0.23  &     7.39 &  1.13 & 
 1.43\\
0    &   638.67 & 0.42  &     1.70 & 0.55 & 0.71  &     1.47 &  0.39 & 
 0.44\\
1    &   475.73 & 0.44  &     2.68 & 0.97 & 0.83  &     2.05 &  0.62 & 
 0.89\\
1   &   503.83 & 0.17  &     1.99 & 0.85 & 0.57  &     5.49 &  1.26 & 
 2.40\\
1    &   535.21 & 0.10  &     0.99 & 0.18 & 0.27  &    11.49 &  2.64 & 
 3.00\\
1   &   567.26 & 0.12  &     0.75 & 0.21 & 0.32  &     5.58 &  1.66 & 
 1.40\\
1    &   591.27 & 0.15  &     0.96 & 0.30 & 0.38  &     6.63 &  1.77 & 
 2.23\\
1   &   625.23 & 0.25  &     1.64 & 0.47 & 0.52  &     2.54 &  0.64 & 
 1.07\\
1    &   660.36 & 0.29  &     1.32 & 0.38 & 0.47  &     1.58 &  0.52 & 
 0.59\\
2    &   367.72 & 0.37  &     0.36 & 0.11 & 0.23  &     3.92 &  0.80 & 
 1.03\\
2   &   396.38 & 0.03  &     0.24 & 0.04 & 0.05  &    26.62 &  4.92 & 
 7.42\\
2  &   426.25 & 0.17  &     0.44 & 0.07 & 0.09  &    17.54 &  3.79 & 
 4.70\\
2    &   455.86 & 0.07  &     0.37 & 0.04 & 0.05  &    43.22 &  6.67 & 
 8.57\\
2    &   485.51 & 0.06  &     0.40 & 0.03 & 0.03  &    57.83 &  6.89 & 
 8.56\\
2    &   515.43 & 0.06  &     0.51 & 0.06 & 0.07  &    33.63 &  4.72 & 
 6.48\\
2   &   545.53 & 0.08  &     0.76 & 0.09 & 0.11  &    12.29 &  1.76 & 
 2.08\\
2   &   575.47 & 0.17  &     1.01 & 0.17 & 0.18  &     4.75 &  0.87 & 
 1.11\\
2   &   605.11 & 0.37  &     1.33 & 0.21 & 0.23  &     2.80 &  0.45 & 
 0.57\\
2   &   636.60 & 0.70  &     1.70 & 0.55 & 0.71  &     0.56 &  0.15 & 
 0.18\\
3  &   379.13 & 0.82  &     0.36 & 0.11 & 0.23  &     0.85 &  0.21 & 
 0.26\\
3    &   408.13 & 0.62  &     0.24 & 0.04 & 0.05  &     5.80 &  1.18 & 
 1.59\\
3    &   436.93 & 0.19  &     0.44 & 0.07 & 0.09  &     3.78 &  0.84 & 
 1.13\\
3   &   467.03 & 0.15  &     0.37 & 0.04 & 0.05  &     9.25 &  1.39 & 
 1.94\\
3   &   496.95 & 0.10  &     0.40 & 0.03 & 0.03  &    12.49 &  1.69 & 
 2.06\\
3    &   526.69 & 0.16  &     0.51 & 0.06 & 0.07  &     7.30 &  1.08 & 
 1.33\\
3    &   556.69 & 0.26  &     0.76 & 0.09 & 0.11  &     2.66 &  0.45 & 
 0.49\\
3    &   587.09 & 0.40  &     1.01 & 0.17 & 0.18  &     1.03 &  0.23 & 
 0.29\\
3    &   618.62 & 0.62  &     1.33 & 0.21 & 0.23  &     0.61 &  0.11 & 
 0.13\\
3   &   650.38 & 0.61  &     1.70 & 0.55 & 0.71  &     0.12 &  0.03 & 
 0.04\\
\hline
\label{tbl-2}
\end{tabular}
\end{center}
\end{table*}

\begin{table*}[!htb]
\begin{center}
\caption{Frequencies, amplitudes, linewidths and their associated uncertainties, for the models described in the text.}
\begin{tabular}{ccccc}
\hline
\hline
Method & M (M$_\odot$) & R (R$_\odot$) & $t$ (Gyr) & $\log g$  \\
\hline
\noalign{\smallskip}
BaSTI Canonical& 1.51$^{+0.12}_{-0.17}$ & 3.14$^{+0.08}_{-0.12}$ & 2.16$^{+1.13}_{-0.46}$ & 3.62$^{+0.01}_{-0.02}$  \\
BaSTI Overshooting& 1.47$^{+0.09}_{-0.12}$ & 3.12$^{+0.06}_{-0.09}$ & 2.50$^{+0.80}_{-0.25}$ & 3.62$^{+0.01}_{-0.01}$  \\
RADIUS & 1.56\,$\pm$\,0.06 & 3.16\,$\pm$\,0.05 & 1.98\,$\pm$\,0.34 & 3.63\,$\pm$\,0.01\\
ASTEC & 1.55 $\pm$ 0.13 & 3.16 $\pm$ 0.18 & 1.87 $\pm$ 0.51 &  3.62 $\pm$ 0.03 \\
\hline
\label{table5}
\end{tabular}
\end{center}
\end{table*}


\section{Stellar modelling}
\label{Sec_Model}
Using the asteroseismic and spectroscopic constraints described in the previous sections, we modeled KIC~8561221 using different methods to gain some knowledge on the internal structure of the star.

An initial set of stellar parameters was determined using a {\it grid-based} method. This technique relies on a pre-constructed grid of stellar tracks or isochrones where the global seismic quantities ($\Delta\nu$ and $\nu_\mathrm{max}$) and atmospheric parameters are used as inputs to determine a best fitting model. This model and the uncertainties on its fundamental parameters such as mass, radius, and age are usually obtained using a maximum likelihood approach coupled with Monte Carlo simulations, as described in \citet{2010ApJ...710.1596B} and \citet{2012ApJ...757...99S}.

We obtained two sets of grid-based modelling results using the BaSTI isochrones \citep{2004ApJ...612..168P} especially computed for asteroseismic analysis, as described in \citet{2013ApJ...769..141S}. One set considered the canonical BaSTI isochrones while the other used a grid of non-canonical isochrones including the effect of overshooting in the main-sequence phase when a convective core is present. A third set of grid-modelling results was obtained using the RADIUS pipeline described in \citet{2009ApJ...700.1589S}, where the stellar properties are estimated from the best fitting model (the one with the lowest $\chi^2$). 1-$\sigma$ uncertainties are estimated as one-sixth of the maximum range of the selected model.

As individual pulsation modes are also available for KIC~8561221, a fourth set of results were determined from a search of the best fitting model which included the frequencies. In this case, the evolutionary tracks were computed using the ASTEC code \citep{2008Ap&SS.316...13C} with the following input physics: OPAL 2005 equation of state \citep{2002ApJ...576.1064R}, OPAL opacity tables \citep{1996ApJ...464..943I} supplemented by low-temperature opacities from \citet{2005ApJ...623..585F}, and NACRE compilation of nuclear reaction rates \citep{1999NuPhA.656....3A}. The \citet{1993oee..conf...14G} solar mixture was adopted. The models did not include microscopic diffusion or overshooting, and convection was treated using the mixing-length theory (MLT) of \citet{1958ZA.....46..108B}. The grid of models covered the range of 1.0-1.6M$_{\odot}$ in mass, 0.24-0.32 in fractional initial helium abundance, and 0.01 to 0.07 in metallicity ($Z/X$),
while the mixing length parameter which defines the mixing length in terms of pressure scale height, $\alpha_{\rm{MLT}}$, was fixed to 1.8.
\begin{figure}[!htb]
\centering
\includegraphics[scale=0.37]{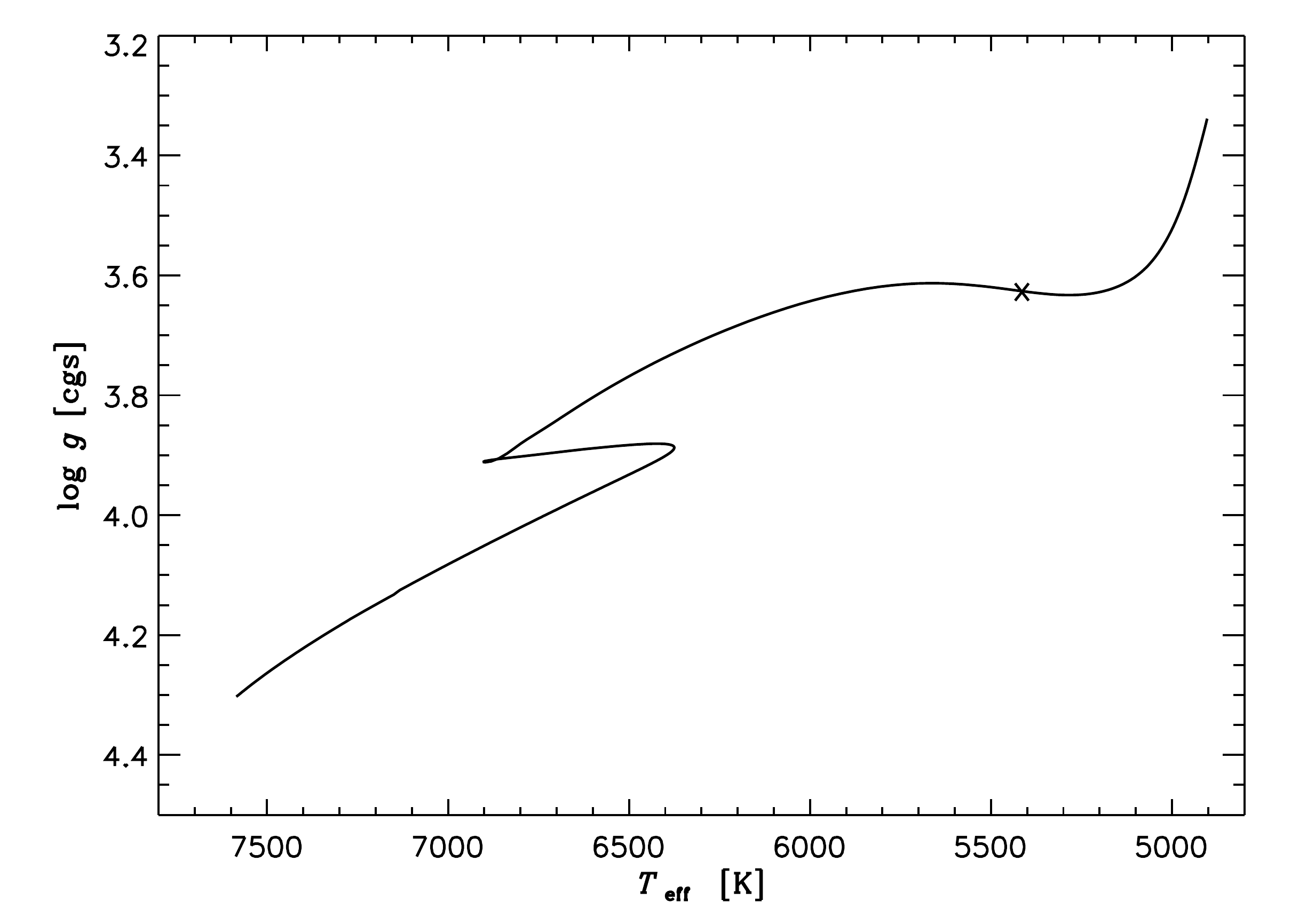}
\caption{HR diagram showing the evolutionary track of KIC~8561221. The cross indicates the location of the best fitting model computed using the ASTEC code.}
\label{fig_HR}
\end{figure}

\begin{figure}[!htb]
\centering
\includegraphics[scale=0.37]{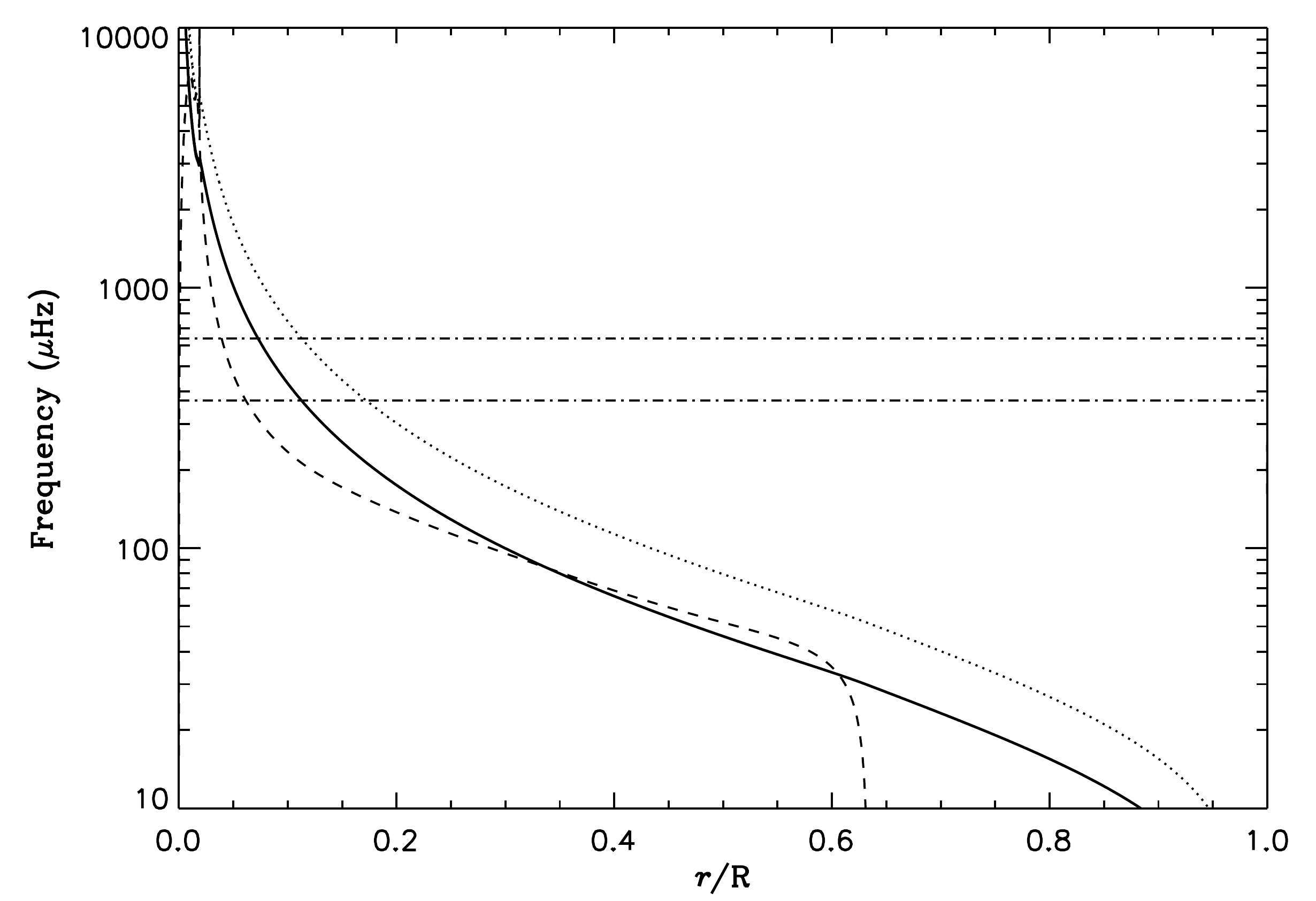}
\caption{Propagation diagram computed using the best fitting ASTEC model. The dashed line represents the Brunt-V\"ais\"al\"a frequency. The continuous and dotted lines are the Lamb frequencies for the $\ell=$1 and 2 modes, respectively. The horizontal dot-dashed lines represent the frequency range in which the modes were detected.}
\label{fig_prop}
\end{figure}

The individual oscillation frequencies of the models were calculated using the Aarhus adiabatic pulsation package \citep[ADIPLS,][]{2008Ap&SS.316..113C}. The empirical near-surface correction suggested by \citet{2008ApJ...683L.175K} was applied to the model frequencies, although additional care was given to the mixed modes following \citet{2013ApJ...763...49D}. Since most of the non-radial modes of this star are mixed modes with high mode inertias relative to those of the radial modes, the magnitude of the correction was divided by the ratio between the mode inertia and the inertia of a radial mode at the same frequency \citep{2010aste.book.....A}. The model presented here (see Fig.~\ref{fig_HR})  was selected based on a $\chi^2$ minimization between the individual observed frequencies and the model frequencies. In Fig.~\ref{fig_prop} we have shown the Brunt-V\"ais\"al\"a  and the Lamb frequencies for the dipole and quadrupole modes, as well as the range in which the modes were observed.

The results from the different methods are given in Table~\ref{table5}. They are in remarkably good agreement (within 1$\sigma$).

\section{Discussion}
\label{Sec_disc}
The observed Fourier spectrum of the $\ell= 1$ modes in KIC~8561221 shows modes with normal amplitudes at high frequencies, while 
at lower frequencies their amplitudes drop at a level close to the noise (see Figs.~ \ref{fftzoom} and \ref{fig_HLA} (C)). 
The threshold frequency is not exactly $\nu_{\rm{max}}$ but close to it. We do not have a clear explanation of the reason for 
this behaviour, but  we can perform computations in order to reject some possible hypothesis.

First the threshold frequency around $\nu_{\rm{max}}$ is not present for most of the other stars with
depressed dipolar modes \citep{2012A&A...537A..30M}. 
Therefore, this circumstance can be a coincidence and the threshold frequency could be related to a perturbation induced by other
phenomena such as a fast rotation or a magnetic field that modifies the oscillation properties.

\subsection{Influence of a fast internal rotation}
The first two stars shown to have depressed dipole modes also display a modulation in the light curve that was interpreted as due to 
star spots (Garc\'\i a et al. 2011, KASC-4 meeting). For both stars, a fast surface rotation rate was inferred. An initial link 
between the reduction in the amplitude of the dipole modes and a fast surface rotation was proposed. Unfortunately, this idea 
was very rapidly abandoned.
Indeed, from the ensemble analysis performed by \citet{2012A&A...537A..30M}, the reduction in amplitude of the dipole modes is 
more pronounced for stars with higher $\nu_{\rm{max}}$. If the physical mechanism depended on stellar rotation, 
we would have expected to find some scatter in the visibilities because the rotation rate of the star is not a direct function of 
$\nu_{\rm{max}}$ in the subgiant or giant evolutionary stage.

Concerning KIC~8561221, our analysis provides us with two indicators of rotation. First, the analysis of the rotational 
modulation of the light curve allows us to retrieve the surface rotation frequency of around 0.13 $\mu$Hz. Moreover, the seismic 
characterisation provides us with the mean rotational splitting for $\ell$=2 and $\ell$=3 modes, that is around 0.29 and 0.65 
$\mu$Hz (see Fig.\ref{dist_rot_ang}). Computing the rotational kernels with a stellar model as described in the previous section,
we found that to explain this difference, the radiative region should --on average-- rotate 4 to 8 times faster than the 
convective zone (assumed to spin at the surface value). 
This is in good agreement with the previously analysed red giant stars that are in a similar evolutionary stage. Indeed, for
the three, more evolved, red giants analysed by \citet{2012Natur.481...55B}, it has been found that their cores rotate 10 times 
faster than the envelope, while for another early red giant star, KIC~7341231, \citet{2012ApJ...756...19D} inferred a 
core-to-surface rotation ratio of five.
As such, KIC~8561221, rotates sufficiently slowly that a first-order perturbative analysis is sufficient to model the effects of 
rotation \citep{2013A&A...549A..75G,2013A&A...554A..80O}.
Nonetheless, in what follows, we will consider higher order effects to see whether they can provide insight into the present 
observations.

At higher rotation rates, one should observe uneven frequency multiplets.  
This is caused by higher order effects of rotation and, in the case of red giants, by different p-g mode trapping of individual 
members of a given frequency multiplet, thereby leading to different mode inertias \citep{2013A&A...554A..80O}. 
If one focuses on $m=0$ modes, centrifugal deformation affects the frequencies of quadrupole modes before those dipole modes 
\citep[this can be seen, for instance, in the frequency spectra of ][]{2013A&A...550A..77R}.  
Hence, in order to modify the frequencies of axisymmetric
dipole modes alone, one could imagine that only the innermost parts of the star
is rotating very rapidly. In this case, only the mixed dipole modes would be
affected as they are the only ones which reach this region while retaining a significant amplitude at the stellar surface. Indeed, the coupling of
the $\ell=2$ modes would be small in stars close to the base of the red-giant branch as KIC 8561221. However, the
ensemble analysis done by \citet{2012A&A...548A..10M} showed that the frequencies of the dipole modes are
located where they are expected from theory. Likewise, the dipole modes measured in KIC
8561221 at high frequency are at the expected frequencies and the only issue
is a smaller-than-expected amplitude. 

One can then turn to mode visibilities. Recently, \citet{2013A&A...550A..77R}
carried out visibility calculations in rapidly rotating $2M_{\odot}$ ZAMS models.  
Their Fig.~7 indicates an increase in the visibilities of dipole modes at $10$ and $20\%$ of the critical rotation rate. 
However, the mode normalisation applied in that work is inappropriate for solar-like oscillators. If we renormalise these modes 
so that their kinetic energy is constant, then at low inclination angles, dipole modes are more visible than other modes in the 
non-rotating case, but then their amplitudes decrease relative to the other modes so as to reach a similar level as the $\ell=0$ modes at $30\%$ of 
the critical rotation rate.  Although this effect goes in the right direction for the present study, it is way too subtle given 
that the surface rotation rate of KIC~8561221 is below $1\%$ of the critical rotation rate, and that the observed amplitudes of 
the $\ell=1$ modes are significantly below that of the $\ell=0$ modes. Hence, rapid rotation 
remains an unlikely explanation to the lower amplitudes of the dipole modes.

Up to now there is no simple explanation that could link the reduction of $\ell$=1 mode amplitudes to a rapid internal rotation. 
To go further, one should study the impact of rotation on the excitation and damping of modes, and that requires a non-adiabatic 
treatment accounting for rotation. 

\begin{figure*}[!htb]
\centering
\includegraphics[scale=0.55]{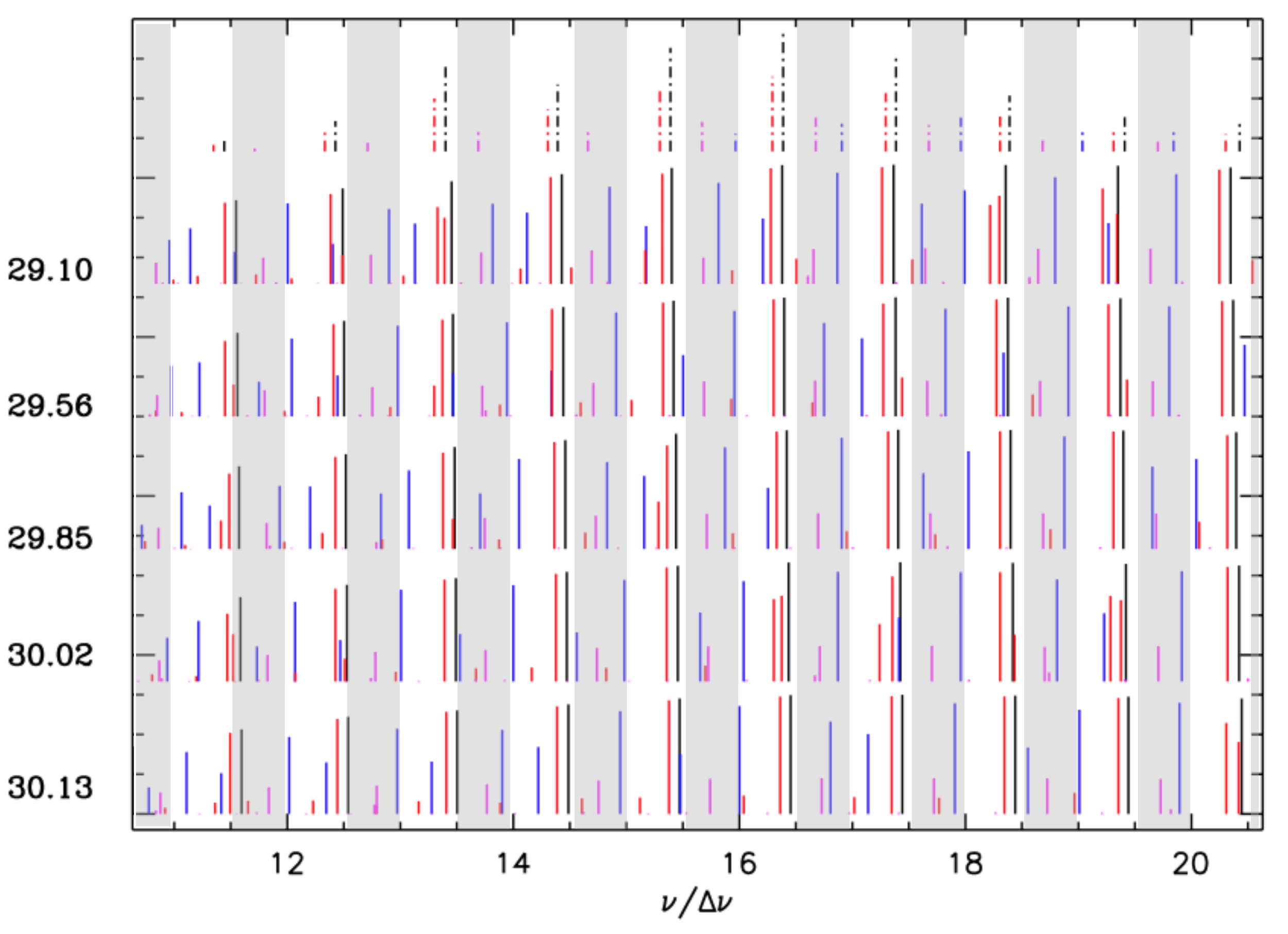}
\caption{Top: Frequency spectrum corresponding to the observed modes (dashed lines). All other spectra (solid lines) are based on theoretical models along a $1.5M_{\odot}$ evolution sequence with large separations $\Delta\nu$ around the observed value. 
Theoretical frequencies have been decreased by a constant factor to improve visual comparison. Different colors corresponds to different degrees as in Fig.~\ref{fig_HLA}: $\ell=0$ (black), $\ell=1$ (blue), $\ell=2$ (red)
and $\ell=3$  (pink). The amplitudes of the theoretical modes have been scaled to $1/\sqrt{E}$ relative to the radial modes except for the $\ell=3$ modes which in addition contain a $0.3$ factor (to improve the visibility). The numbers on the left are the values of the large separation $\Delta\nu$ in $\mu$Hz. Grey shaded regions are done to guide the eye.}
\label{fig_spec}
\end{figure*}

\subsection{Mode inertia}

Since KIC~8561221 is the least evolved red giant star with depressed dipolar modes we may even ask ourselves whether it belongs
to this sequence of stars \citep{2012A&A...537A..30M}. It is the first star in which we observe both, normal and depressed dipolar modes. In fact,
KIC~8561221 has evolved to a point where $\ell$=1 and 2 modes are mixed, although the amplitudes of the g-dominated $\ell$=2 modes are very low. In particular, between two consecutive radial modes there may not be any $\ell=1$ modes with a clear p-mode character.
This gives rise to the question of whether the depressed dipolar modes in this star are due to a fortuitous circumstance where all
the  $\ell=1$ modes in the lower part of the observed frequency range behave like g-dominated mixed modes with low amplitudes, while regular p modes would be present at higher frequencies.
Fig.~\ref{fig_spec} shows schematic frequency spectra for the observations (top row, with dot-dashed lines) and some models with 
similar global seismic properties. The
models were computed with the CESAM code \citep{1997A&AS..124..597M}, while frequencies were computed in the adiabatic
approximation with the ADIPLS code \citep{2008Ap&SS.316..113C}. Because of the surface effects, the theoretical mode frequencies
would have appeared shifted if they were directly compared with the observed values. To simplify the comparison, the theoretical
frequencies have been decreased by a dimensionless factor to improve visual comparison. Note that although the true correction should be lower for core trapped modes, as indicated in the Sect. 3, in Fig. \ref{fig_spec} we have kept the actual theoretical spectrum
patterns.

The ratio between the surface rms amplitude and the energy of the mode is proportional to $1/\sqrt{E}$ where $E$ is the
dimensionless mode inertia. For low-degree modes excited stochastically in the upper layers of the convection zone one
expect that their energy and damping would be mainly a function of frequency and the observed $m$-averaged amplitude would be of the form $f(\omega)/\sqrt{E_{nl}}$. Hence, in a given frequency range, the ratio between the mean amplitude of the
dipolar and the radial modes would be proportional to $\langle\sqrt{E_{n0}}\rangle/\langle\sqrt{E_{n1}}\rangle$.
In the absence of mixed modes this ratio is roughly constant. The same behaviour can be expected
for evolved red giants where p-dominated mixed $\ell=1$ modes would contribute the most to $\langle\sqrt{E_{n1}}\rangle$.
In fact, from an observational point of view,
\citet{2012A&A...537A..30M} computed the integrated power spectrum over frequency bands centred
around the expected central frequencies for different degrees and found that the power ratios, both
between the $\ell$=1 and $\ell$=0 modes and the $\ell$=2 and $\ell$=0 modes, were approximately constant for all the
``normal'' stars or at most that the ratios change gradually with $\nu_{\rm max}$.
The set included RGB stars with $\nu_{\rm{max}}<200$ $\mu$Hz, which all have large numbers of mixed $\ell=1$ modes clustered around each  p mode.

\begin{figure}
\centering
\includegraphics[scale=0.38]{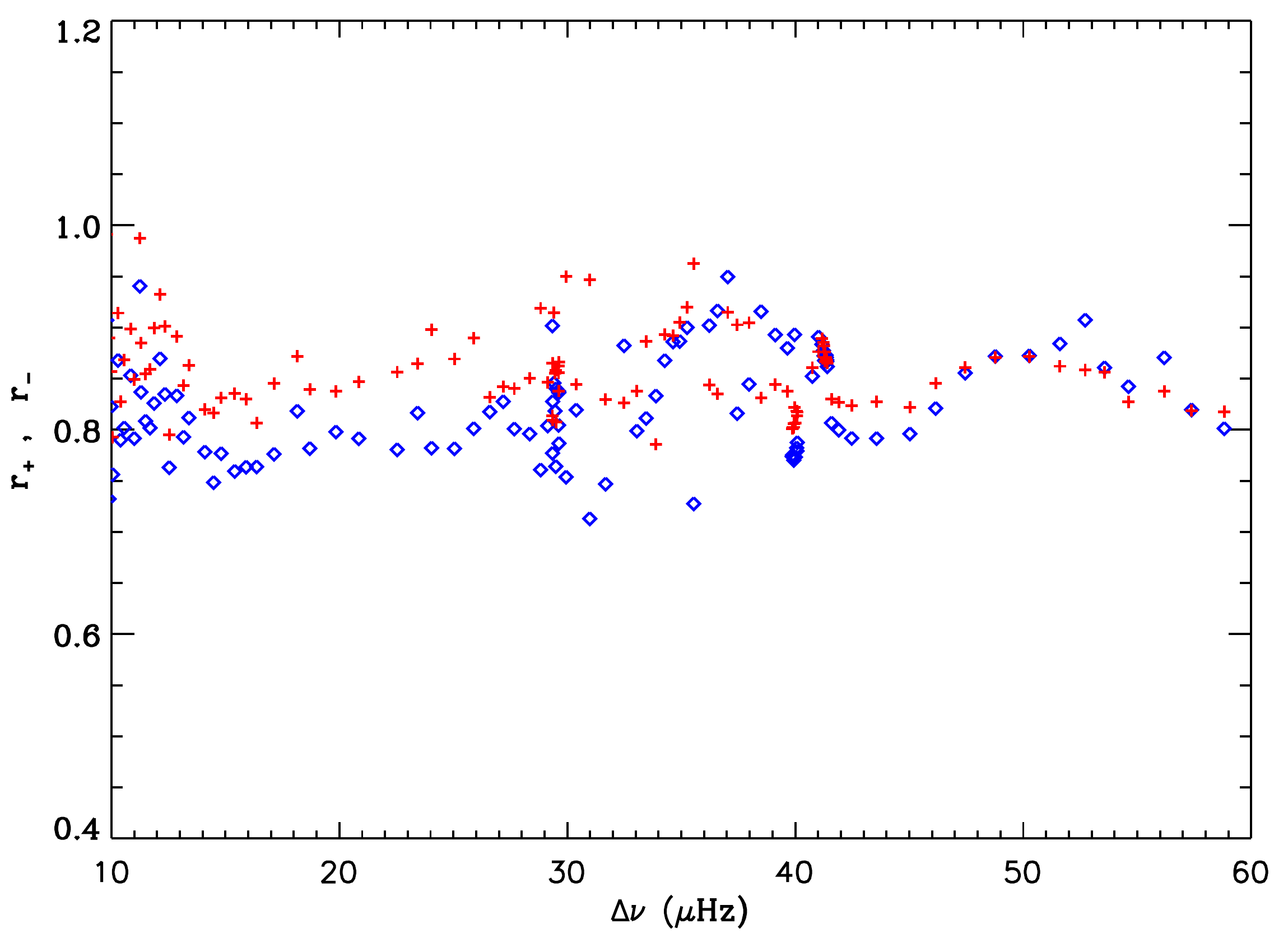}
\caption{Energy ratios $r_+$ (red plusses) and $r_-$ (blue diamonds) versus $\Delta\nu$ computed for a
$1.5M_{\odot}$ evolution sequence. See text for details.}
\label{fig_ratio}
\end{figure}

As mentioned above, KIC~8561221 is in an evolutionary stage in which such arguments could be invalid.
A priori, significative changes in the ratio $\langle\sqrt{E_{n0}}\rangle/\langle\sqrt{E_{n1}}\rangle$ cannot be rejected.
This could explain the observed power spectrum while keeping the theoretical energy ratios.
According to the observed amplitudes in KIC~8561221 the transition between the normal and the depressed dipole modes is around
$\nu_{\mathrm{max}}$. Therefore, we have verified if for some specific models the power above and
below $\nu_{\mathrm{max}}$ is very different in comparison to the normal case.
To be more specific, we estimate $\nu_{\mathrm{max}}$ from the $T_{\mathrm{eff}}$ and $\log g$ computed from the
models and then we search for the radial mode closest to that frequency, $\nu(\ell=0,n=n_{\rm{max}})$.
Finally, we compute the ratios
\begin{equation}
r_{\pm}= \frac{ E( n=n_{\rm{max}}\pm \Delta n/2; \ell=0)} {\left(\sum_i 1/E_i(\ell=1) \right)^{-1} (\Delta n+1/2)}
\sim \frac{\langle A^2 \rangle_{\ell=1}}{\langle A^2 \rangle_{\ell=0}}
\end{equation}
where the sum over $i$ is for all the $\ell=1$ modes in the frequency range
$[\nu(\ell=0, n=n_{\rm{max}}-\Delta n), \nu(\ell=0, n=n_{\rm{max}})]$ for the ``$-$'' solution and in the frequency range
$[\nu(\ell=0, n=n_{\rm{max}}), \nu(\ell=0, n=n_{\rm{max}}+\Delta n)]$ for the ``$+$''  solution.
These ratios are shown in Fig. \ref{fig_ratio} where a value of $\Delta n=3$ has been chosen as a representative example.
In this figure, every pair of red and blue points corresponds to a given model from a
$1.5M_{\odot}$ evolution sequence with the large separation indicated on the ordinate.
As explained above, the observations suggest that this ratio can be interpreted as a power amplitude ratio,
$\langle A^2 \rangle_{\ell=1}/\langle A^2\rangle_{\ell=0}$, except for a constant visibility factor.

As seen in Fig.~\ref{fig_ratio} the power of the $\ell=1$ modes on the left of $\nu_{\mathrm{max}}$ is not very different to
that on the right, despite the number and nature of the $\ell=1$ involved.
In fact, for the models with the highest $\Delta \nu$, most of the $\ell=1$ are p modes whereas for the models with
the lowest $\Delta \nu$, mixed modes are predominant. Although for stars with $\nu_{\mathrm{max}}\sim 30-40$ $\mu$Hz there is
a little higher scatter, only a change of $10\%- 20\%$ in the relative amplitudes
could be explained by the fluctuations in the mean mode inertia, far from the observed situation for KIC~~8561221.
We checked this result with other evolutionary sequences as well as for models with different stellar parameters. The result
is always comparable to the one showed in Fig.~\ref{fig_ratio}.
Hence, we do not have any reason to think that the nature of the mixed modes in the relatively unevolved red giant
KIC~8561221 can make this star different compared to more evolved ones with depressed dipolar modes.

\subsection{Damping}
For common red giants observed by {\it Kepler}, the power of the g-dominated mixed modes have lower width than the regular
p modes. This result can be expected because most of the damping is produced in the envelope layers, while for g-dominated mixed modes this region contribute relatively less to the total work than for p-dominated ones.
However, as shown in Fig.~\ref{fig_HLA}, the width of the dipole modes below $\nu_{\rm max}$ is larger than for the rest of the
modes. A strong damping in the core could explain the lower power observed for these dipole modes.

It is interesting to notice that the damping in the core
is very important for highly evolved --very luminous-- red giants \citep[e.g.][]{1977AcA....27...95D,2012A&A...539A..83D}.
In fact, for these stars the damping is expected to make g-dominated
mixed modes hardly observable. However, $\ell=1$ p-dominated modes are
very effectively trapped in the envelope and hence, at least  one of such modes should have high amplitudes between every radial mode. \citet{2009A&A...506...57D} computed theoretical amplitudes for selected red giants and showed how the damping changes with evolution and luminosity through the red giant branch.  For low-luminosity stars,
the core damping becomes lower than the one in the envelope and the g-dominated mixed modes are then expected to be observed, as is the case for the large majority of red giant studied with CoRoT and {\it Kepler}.

Here we shall consider models that match --to a great accuracy-- the fundamental parameters (and oscillation frequencies) of KIC~8561221 derived
in the previous section, namely the model in Fig.~\ref{fig_spec} with $\Delta\nu$=29.85 $\mu$Hz. This model has
a mass of $M=1.5M_{\odot}$, a luminosity of $L=7.23 L_{\odot}$, and an effective temperature of $T_{\rm{eff}}=5281\,$K.

\begin{figure}[!htb]
\centering
\includegraphics[scale=0.37]{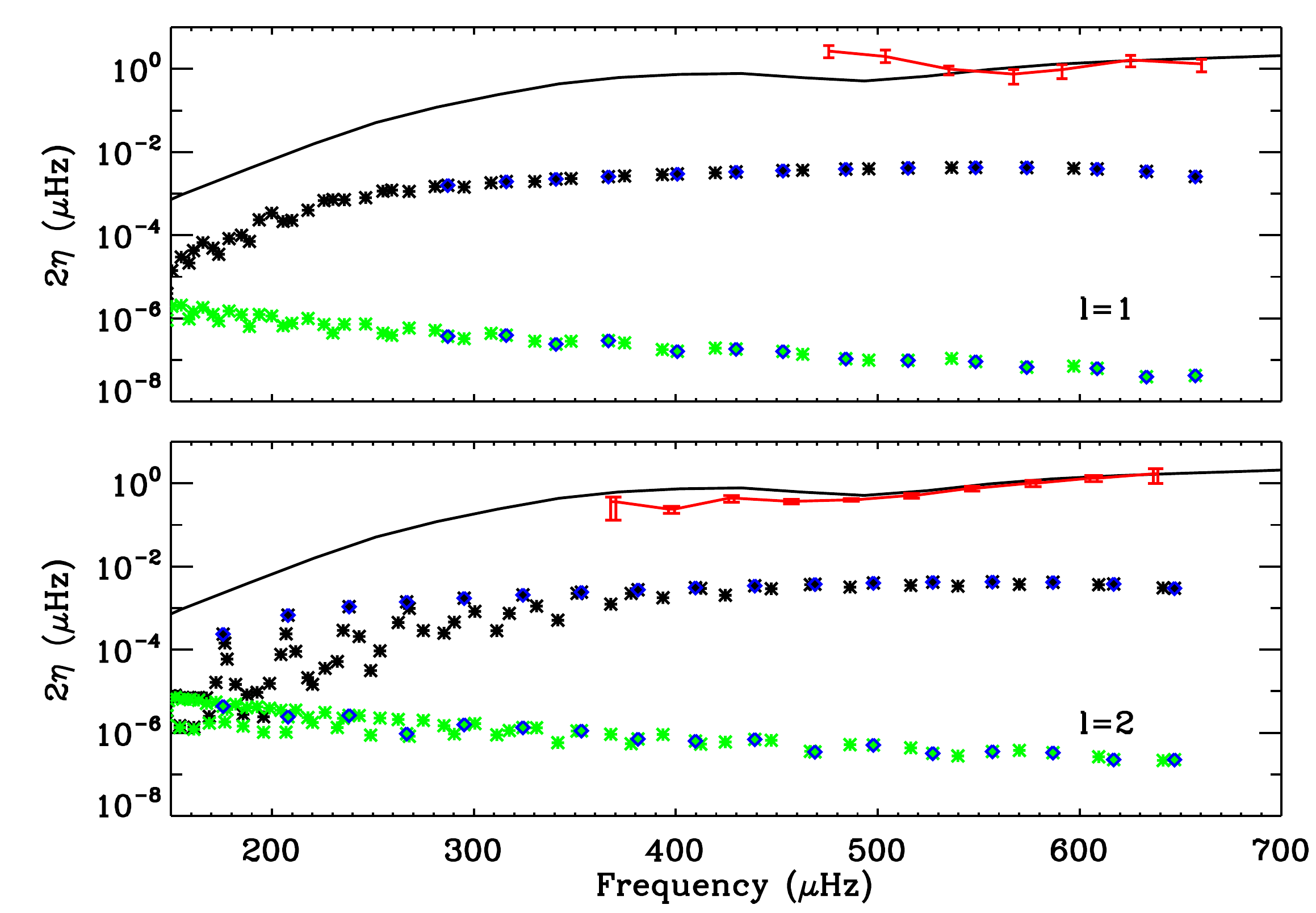}
\caption{Observed and theoretical linewidths for $\ell$=1 modes (top) and $\ell$=0 and 2 modes (bottom). The red lines with error bars correspond to the observational linewidths shown in Fig.~\ref{fig_HLA}.
The solid black lines in both panels are the theoretical estimates for radial modes, obtained from model calculations in which the 
turbulent pressure was included consistently in both the equilibrium structure and stability computations.
For nonradial oscillations only radiative damping was considered.
Black points correspond to the damping rate in the whole star, while green points are for the $g$-mode cavity. 
Blue diamonds are p-dominated mixed modes.}
\label{fig_damping}
\end{figure}

First we estimated pulsation linewidths for radial modes including a full nonadiabatic treatment and
convection dynamics in the stability computations.
Both the convective heat and momentum (turbulent pressure) fluxes were treated consistently in both the equilibrium and pulsation
computations, using the nonlocal generalization of \citet{1977LNP....71...15G,1977ApJ...214..196G} time-dependent convection
model. The computations were carried out as  described by \citet{1999A&A...351..582H} and \citet{2002MNRAS.336L..65H}. For the
nonlocal convection parameters we adopted the values $a^2=b^2=900$.
For the anisotropy parameters $\Phi$, as defined in \citet{2002MNRAS.336L..65H}, the value 1.40 was adopted.
The black continuous lines in Fig. \ref{fig_damping} correspond to
twice the estimated linear damping rates, which are approximately equal to the pulsation linewidths.
The outcome shows a similar frequency dependence of the radial linear damping
rates (linewidths) than what we measure in the Sun, but with values of about 30$\%$ smaller than in the solar case.
For an easier comparison with the observed values we also shown in Fig.~\ref{fig_damping} the observational linewidths given in
Fig. \ref{fig_HLA}.
The theoretical computations, limited to radial modes, reproduce the main features of the observational results for the
modes with regular amplitudes, that is all the radial and quadrupole modes and the highest dipole modes.

In addition, damping rates for nonradial oscillations were computed in the quasi-adiabatic approximation, following
\citet{1989nos..book.....U}. We have included the
$\varepsilon$ excitation although it is negligible. On the other hand, for these nonradial modes we have not included
the dynamics of the convection which would give a damping rate larger than the one considered here. However, with standard
assumptions, the damping should be located very close to the surface and hence the corresponding term would be a function of frequency
alone. Therefore, it can be obtained by extrapolating the results for the radial oscillations.

Fig. \ref{fig_damping} shows these linewidths, that are two times the imaginary part of the frequency, $2\eta$,
against the mode frequency for the indicated model.
Black points correspond to the damping rate computed by considering the whole star.
Modes with a superimposed blue diamond indicates a p-dominated character, that is, with a mode inertia close
to the one that would have a radial mode with the same frequency.
Green points correspond to damping rates computed using only the eigenfunctions in the $g$-mode
cavity. Here blue diamonds are also included to show a p-dominated character.
Although the core (g-mode cavity) damping rate increases rapidly at low frequencies (the scale is logarithmic) it is
negligible compared to the envelope damping through all the observed frequency range.
As a consequence, damping rates for the $\ell=1$ and the $\ell=2$ modes are basically the same at a given frequency
regardless of the degree of mixed character in the mode. What we found here is the situation expected for a normal
star with the same parameters as KIC~8561221 where the $\ell=1$ modes have normal amplitudes.
We have done the same computations for several evolution sequences and
found that for the models with a large separation close to that of KIC~8561221 the results were always similar.

In conclusion, using standard models with the parameters of KIC~8561221, we are able to fit the observed frequencies,
including the $\ell=1$ modes but not the linewidth of the dipole modes that are expected to be comparable with those of the
radial and quadrupole modes.
Hence, if the linewidths were to be interpreted as damping rates, another source of energy dissipation, not included in our
stability computations, must be responsible for the depressed $l$=1 modes.

Since there are evidences of a surface magnetic activity in this star, 
it can be thought that a magnetic dissipation could cause the extra damping.
There are some works dealing with this phenomena, most of
them done in the context of rapidly oscillating Ap stars
\citep[e.g.][]{1983MNRAS.205.1171R,2005JApA...26..213C},
where the energy lost is caused by the dissipation of the low component decoupled from the acoustic waves
after a surface reflection. However such effect would result in very similar damping for all low-degree modes.
Even the presence of a magnetic dynamo, with a strong magnetic field located at the base of the convection zone (i.e. at
about $0.6R$, see Fig. \ref{fig_prop}) would not make any difference. 

\subsection{Influence of a deep magnetic field}

The presence of a deep (fossil) magnetic field in the stably stratified inner core of the star could be evoked as a possible reason for the
reduction of the amplitude of the dipole $\ell$=1 modes. 
Here again, one can expect that the mixed dipole modes would feel the effects of such a magnetic field most sensitively
as they are the only observed modes that can have high amplitudes in the core.

Magnetic fields could contribute to reducing the amplitude of the dipole modes through the following effects:
\begin{enumerate}
\item If there exists a strong (axisymmetric) magnetic field that is not aligned with the rotation axis, the effects of the magnetic field and rotation lift the degeneracy of dipole modes (with a given radial order) with respect to the azimuthal order, $m$, which produces three eigenmodes (for $m=0, \pm$ 1)  with different frequencies in the rotating frame.  If we
observe these modes in the inertial (observer's) frame, each of them has three frequency components due to the advection effects \citep[e.g.][]{GooTho1992,1993PASJ...45..617S}. Therefore, it is possible to observe nine frequency components in total
that originate from dipole modes with a given radial order. Provided that the kinetic energy of each eigenmode is not changed by the magnetic field, the increase in the number of the frequency components results in the reduction of their amplitudes.

\item More energy is needed to displace each mass element to a fixed amplitude against the Lorentz force in the presence of the magnetic field. This means that the amplitude is lower, if the mode energy is the same.
\item The magnetic field generally modifies the eigenfunction of the dipole modes to have components with higher spherical degrees, which suffer stronger geometrical cancellation when observed.  Since part of the mode energy goes to those additional components with lower visibility, the total amplitude of the distorted dipole mode in the power spectrum gets
smaller.
\end{enumerate}

It is outside the scope of the present work to quantify the magnitude and geometrical topology of the magnetic field to reproduce the observed effect on the $\ell=1$ modes while keeping, at the same time,  a negligible influence of this magnetic field on higher degree modes. However,
there is no  evidence of magnetic splitting or shift in other modes. The standard models fits the frequencies of the $\ell=$0, 2, and $3$ modes with a similar accuracy to those of normal red giants. Therefore, there is no observational support in favour of a strong (fossil) magnetic field in the core of this star. 

\section{Conclusions}

In this paper we have studied KIC~8561221, which is the least evolved observed star with depressed dipole modes, in order to better understand the possible physical mechanisms responsible for the reduction of the amplitude of these modes.

KIC~8561221 background parameters were found to be normal and they follow the relation described in
\citet{2011ApJ...741..119M}. The detailed analysis of the light curve showed a modulation in the brightness of the star
that we assumed to be due to spots crossing the stellar disk. Our analysis of the surface rotation  suggests a rotation period of 91 days.
 It is slower than expected following standard calculations for stars in the range 1.4 to 1.5 $M_\odot$. A lower rotation rate is possible if the precursor was  a chemically peculiar star or if the star slowed down due to a strong magnetic braking. Indeed, the variation in the scale-averaged time series around the rotation period indicate that the star still has a surface dynamo magnetic field. We have also unveiled changes in the photometry suggesting an on-going magnetic activity cycle.

The detailed seismic analysis concluded that KIC~8561221 is an early red giant located at the base of the RGB.
Modes with $\ell$=0, 1, 2, and 3 were observed and seismically characterized. Frequencies of all the detected modes were found at
the expected locations, ruling out a very fast spinning interior. 
Moreover, the splitting of the modes with $\ell=2, 3$ indicate that the interior is rotating 4 to 8 times faster than the surface,
a similar rate than other RGB stars.

However, the dipole amplitudes were found to be much lower than expected, in particular below $\nu_{\mathrm{max}}$ , while their widths were higher than expected. The amplitudes of the $\ell= 2$
modes were also found to be lower than expected, although to a
lesser extent. While the latter could be explained by an energy leakage towards the g-dominated $\ell=2$ mixed modes, current models of mode damping are unable to explain the observed reduction in the amplitudes of the dipole modes. Contrary to more evolved stars in the sequence,
in KIC~8561221 only one or two dipole modes are expected between two consecutive radial modes. This means that for some radial
orders there is not a single clear p-dominated $\ell=1$ mode. However, our analysis of the mode inertia shows that even when this happens it cannot explain the observed amplitude of the dipole modes. Hence, we expect that KIC~8561221 has
the same unknown physical mechanism operating as in the other stars showing low-amplitude dipole modes. 
As shown in this paper, there are evidences for a strong dynamo magnetic field at the surface of KIC~8561221.
It is very tentative to link this surface magnetic activity with the amplitude reduction of the dipolar modes. 
However, it seems that only a fossil magnetic field in the core could differentiate between the $\ell=1$ modes and other non-radial
oscillations. The null observational evidence of magnetically split components or frequency shift makes the 
hypothesis of a strong core magnetic field unlikely. 

The next step will be to do an ensemble study of all the {\it Kepler} red giants showing depressed dipole modes, not only in the set of stars analysed by \citet{2012A&A...537A..30M} but also in the 13,000 red-giant stars studied by \citet{2013ApJ...765L..41S}. Complementary ground-based observations of some of these stars could also be necessary to solve this puzzle.

\begin{acknowledgements} 
The authors wish to thank the entire {\it Kepler} team, without whom these results would not be possible. The authors thank Dr. D. Huber for useful comments and discussions as well as Dr. Y. Elsworth, S. Hekker and H. Kjeldsen for the coordination activities inside KASC. Funding for this Discovery mission is provided by NASAÕs Science Mission Directorate. We have used data obtained by the NARVAL spectrograph mounted at the T\'elescope Bernard Lyot (USR5026) operated by the Observatoire Midi-Pyr\'en\'ees, Universit\'e de Toulouse (Paul Sabatier), Centre National de la Recherche Scientifique of France.
We also thank all funding councils and agencies that have supported the activities of KASC Working Group 1. Authors acknowledges the KITP staff of UCSB for their hospitality during the research program ÒAsteroseismology in the Space AgeÓ. This research was supported in part by the National Science Foundation under Grant No. NSF PHY05-51164. TC, GRD, RAG, and SM have received funding from the European CommunityÕs Seventh Framework Program (FP7/2007-2013) under grant agreement no. 269194 (IRSES/ASK). DRR is financially supported through a postdoctoral
fellowship from the ``Subside f{\'e}d{\'e}ral pour la recherche 2012'', Universit{\'e} de Li{\`e}ge,  which is gratefully acknowledged. Funding for the Stellar Astrophysics Centre is provided by The Danish National Research Foundation (Grant DNRF106). The research is supported by the ASTERISK project (ASTERoseismic Investigations with SONG and Kepler) funded by the European Research Council (Grant agreement no.: 267864). NCAR is supported by the National Science Foundation. This research was supported in part by the Spanish National Research Plan under project AYA2010-17803. This work partially used data analysed under the NASA grant NNX12AE17G. 
\end{acknowledgements} 
\newpage

\begin{appendix} 
\section{Evaluating the reliability of the fitted results}
\label{Annex}

In order to verify the reliability of the results, we tested the robustness of our measurements of the frequencies, widths, heights and rotational splittings using simulated spectra that reproduce the general properties of KIC 8561221. The present appendix  summarises the results from these tests. 


Usually, reliability is assessed using artificial spectra that mimic the known properties of the stellar oscillations and of the noise by the means of a limit spectrum and of an excitation function. However, this assumes that the physics and the noise properties are \emph{perfectly} known. This is not the case as stars often have a different behavior than the one predicted by theory (e.g. the case of the so-called surface effects or, in the present case, puzzling depressed dipole modes). Therefore, rather than an artificial spectra, we used a modified spectrum of a reference star (KIC~9574283) with very similar global properties (cf. Table \ref{tbl-3}), but with clear split $\ell=1$ modes. 

\begin{table}[!hb]
\begin{center}
\caption{Global properties of the modes for KIC 8561221, compared to KIC 9574283 (reference star), using a MCMC mode fitting algorithm. }
\begin{tabular}{ccc}
\hline
\hline
												 & KIC 8561221								& KIC 9574283 \\
\hline												 
$\Delta\nu$ ($\mu$Hz)      & $29.79 \pm 0.14$						&  $29.8 \pm 0.21$ \\
$\nu_{\rm{max}}$ ($\mu$Hz)      & $491 \pm 5$						&  $\approx 450$ \\
$\Delta P_{1}$ (sec)   & \text{Unknown}						& $ \approx 117$ \\
Maximum SNR ($\ell=0$)     & $\approx 30$						&  $\approx 35$ \\
Inclination (deg)					 & $\leq 30$ 						&  $44.8 \pm 1.4$ \\
$\delta\nu_s(\ell=1)$						 & \text{Unknown}						&  \text{varies from 0.1 to 0.75} \\
$\bar{\delta\nu_s(\ell=2)}$			 & $[0.3 , 0.6]$						&  $0.15 \pm 0.008$ \\
$\bar{\delta\nu_s(\ell=3)}$			 & $[0.3 , 0.6]$						&  $0.097 \pm 0.015$ \\
\hline
\label{tbl-3}
\end{tabular}
\end{center}
\end{table}

Physics of stochastically driven modes tells us that a real star power spectrum $S_R^{*}(\nu)$ can be written as:
\begin{equation}
	S_R^{*}(\nu) = M_R^{*}(\nu) F(\nu)  \; ,
\end{equation} 
where $M_R^{*}$ is the limit spectrum and $F(\nu)$ is a random noise function \citep[see more details in][]{1998A&A...333..362F}. $M_R^{*}(\nu)$ is unknown by definition but can be written as a function of an approximate model $M_R(\nu)$ and of an error function $\epsilon_R(\nu)$. Therefore, $M_R^{*}(\nu) = M_R(\nu) + \epsilon_R(\nu)$ and we can define a new noise function $F'(\nu)$ as follows:
\begin{equation}
	F'(\nu) \equiv \frac{S_R^{*}(\nu)}{ M_R(\nu)} = F(\nu) \left( 1 + \frac{\epsilon_R(\nu)}{ M_R(\nu)} \right) \;.
\end{equation}
In such case, $F'(\nu)$ contains information on the excitation function, the instrumental noise, and the model inaccuracies. Using $F'(\nu)$ we can now build a simulated spectrum $S_A$, where:
\begin{equation}
	S_A(\nu) = M_A(\nu) F(\nu) \left( 1 + \frac{\epsilon_R(\nu)}{ M_R(\nu)} \right) \;.
\end{equation}
Here, $M_A(\nu)$ is the desired limit spectrum and corresponds to an altered version of the model $M_R(\nu)$ such that $M_A(\nu)=\alpha(\nu) M_R(\nu)$. In this expression, $\alpha(\nu)$ acts as a filter.

\begin{figure}[!htbp]
\begin{center}
	\includegraphics[width=9cm, angle=180,trim=1.8cm 2cm 4cm 1.8cm ]{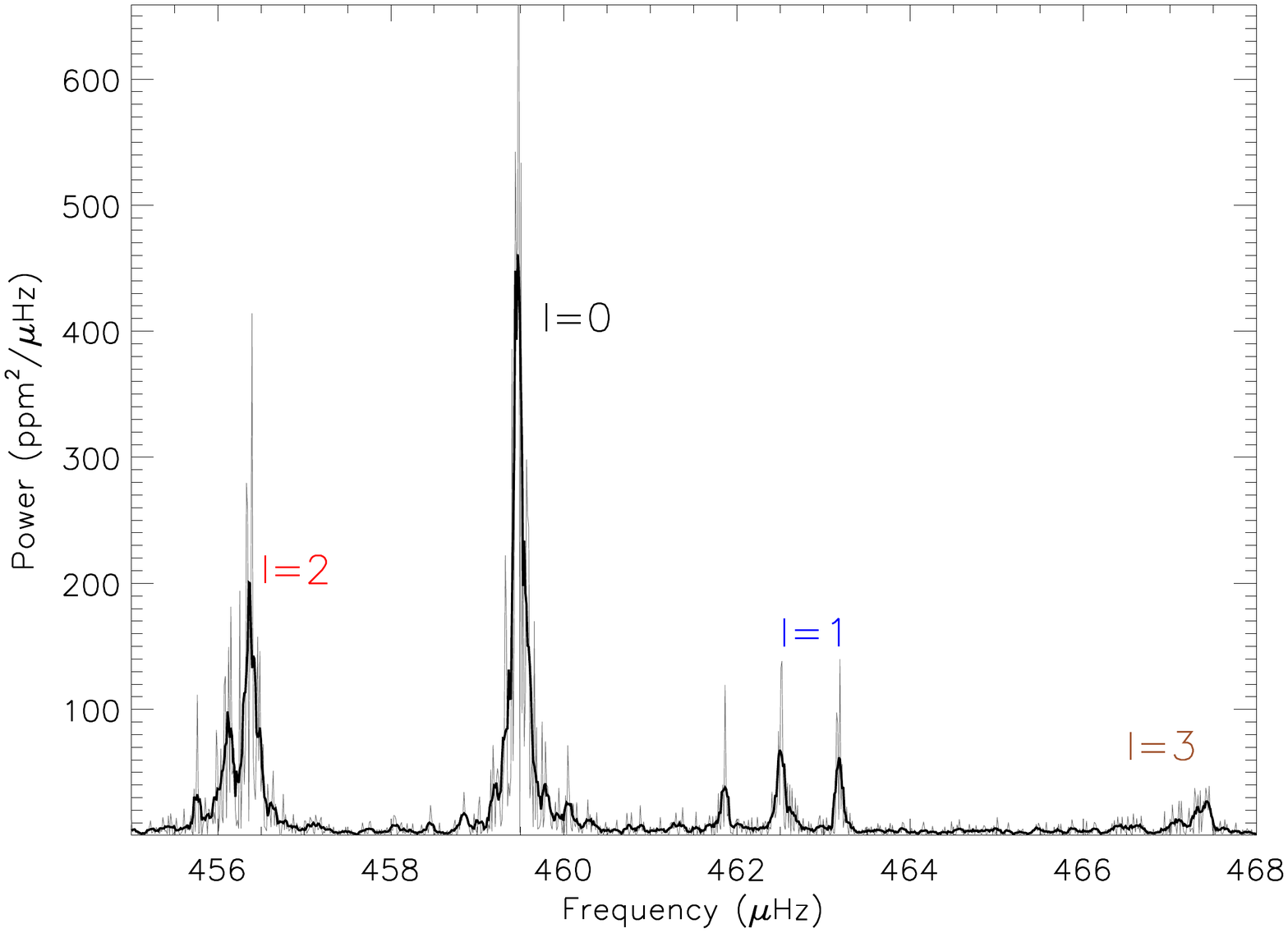}
	\includegraphics[width=9cm, angle=180,trim=1.8cm 2cm 4cm 1.8cm ]{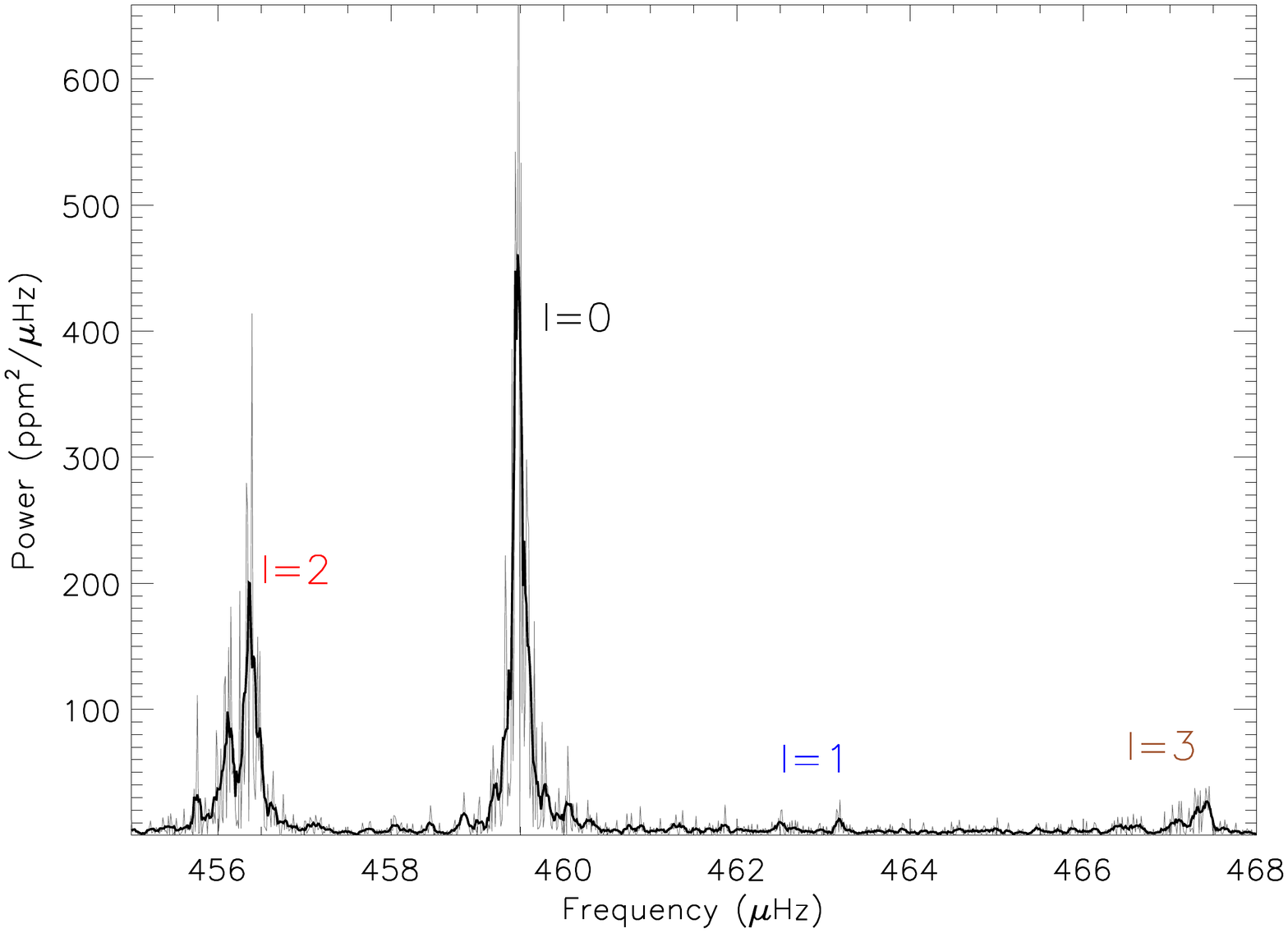}
	\caption{Simulated spectra generated by modifying the signal-to-noise of dipole modes of the reference star KIC~9574283. Top: the original reference spectrum. Bottom: the modified spectrum with a SNR of the dipole modes $\approx 2$, such as in  KIC~8561221.}
\label{Fig:Annex-1}
\end{center}
\end{figure}

In our application, $M_R(\nu)$ is determined by the best fit of the reference power spectrum of the reference star. We modified $\alpha(\nu)$ in order to reduce the height of the $\ell=1$ modes to a similar signal-to-noise as in KIC~8561221 (SNR $\approx 2$) when their frequency is greater than $\nu_{\rm{max}}$, and by removing the $\ell=1$ modes when their frequencies are lower than $\nu_{\rm{max}}$. All the other mode parameters were unchanged. An example of the results can be seen in Fig.~\ref{Fig:Annex-1}.

This method ensures that the simulated spectra contains the unknown physics from the reference star as well as any unknown source of noise, which makes it very realistic. In addition, the 'alteration factor' $\alpha(\nu)$ is a known deterministic modification that only controls the limit spectrum properties. Therefore, by modifying $\alpha(\nu)$, we are sure that we evaluate only its effect and not, for example, an effect from the random noise. In our case, we can therefore evaluate the effect of low signal-to-noise $\ell=1$ modes on parameters of the reference star KIC~9574283, by comparing the results of the original spectrum of this star with the modified spectrum.  This method has also been applied to a second star, KIC~8751420, with a higher inclination angle ($64.4 \pm 1.5$) and a higher $\ell=0$ signal-to-noise ($\approx 55$). The conclusion being the same, we only describe the results on KIC~9574283.

\begin{figure*}[!htbp]
\begin{center}
	\includegraphics[width=7cm,trim=8cm 2cm 2cm 0cm]{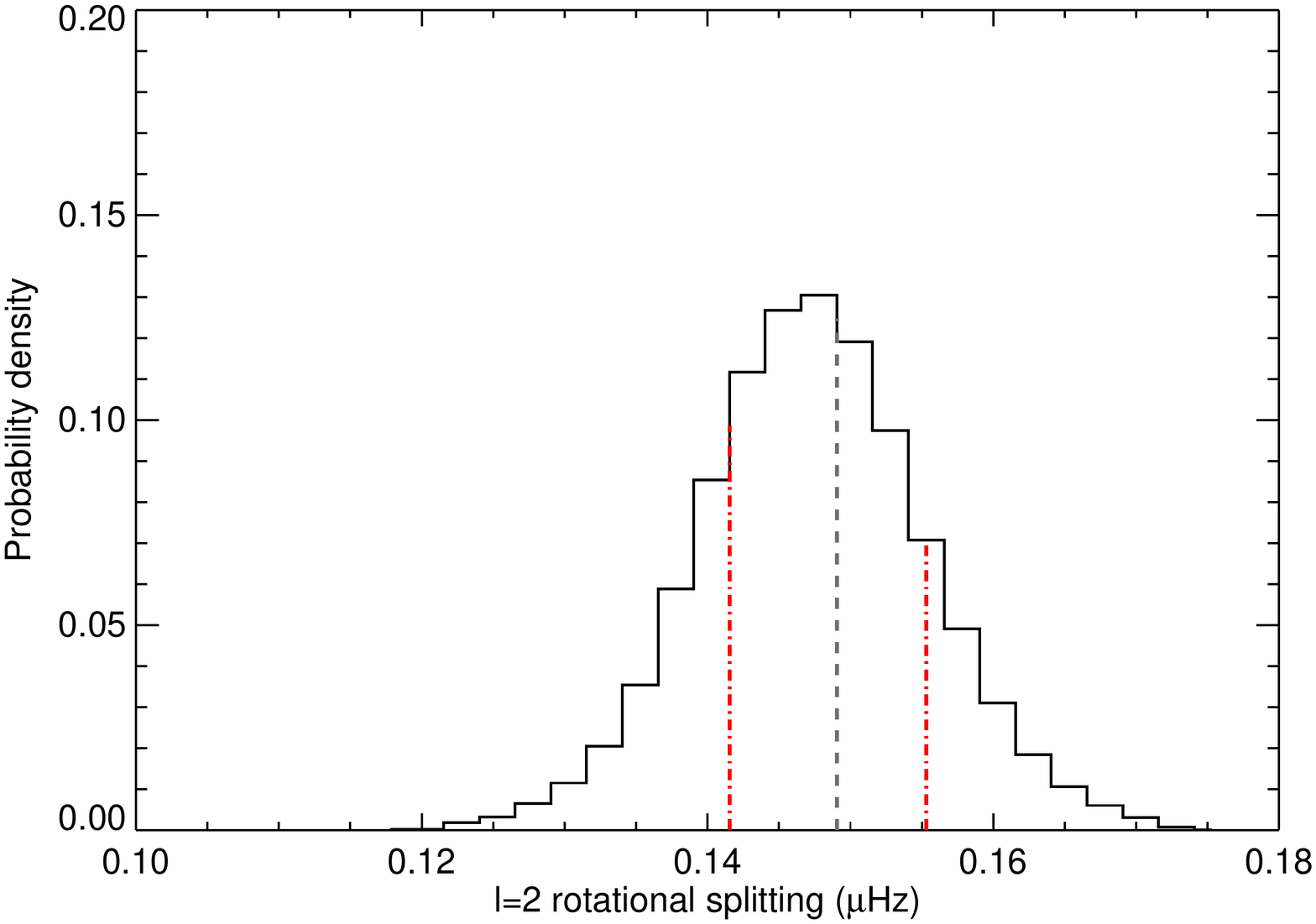}
	\includegraphics[width=7cm,trim=3cm 2cm 7cm 2cm]{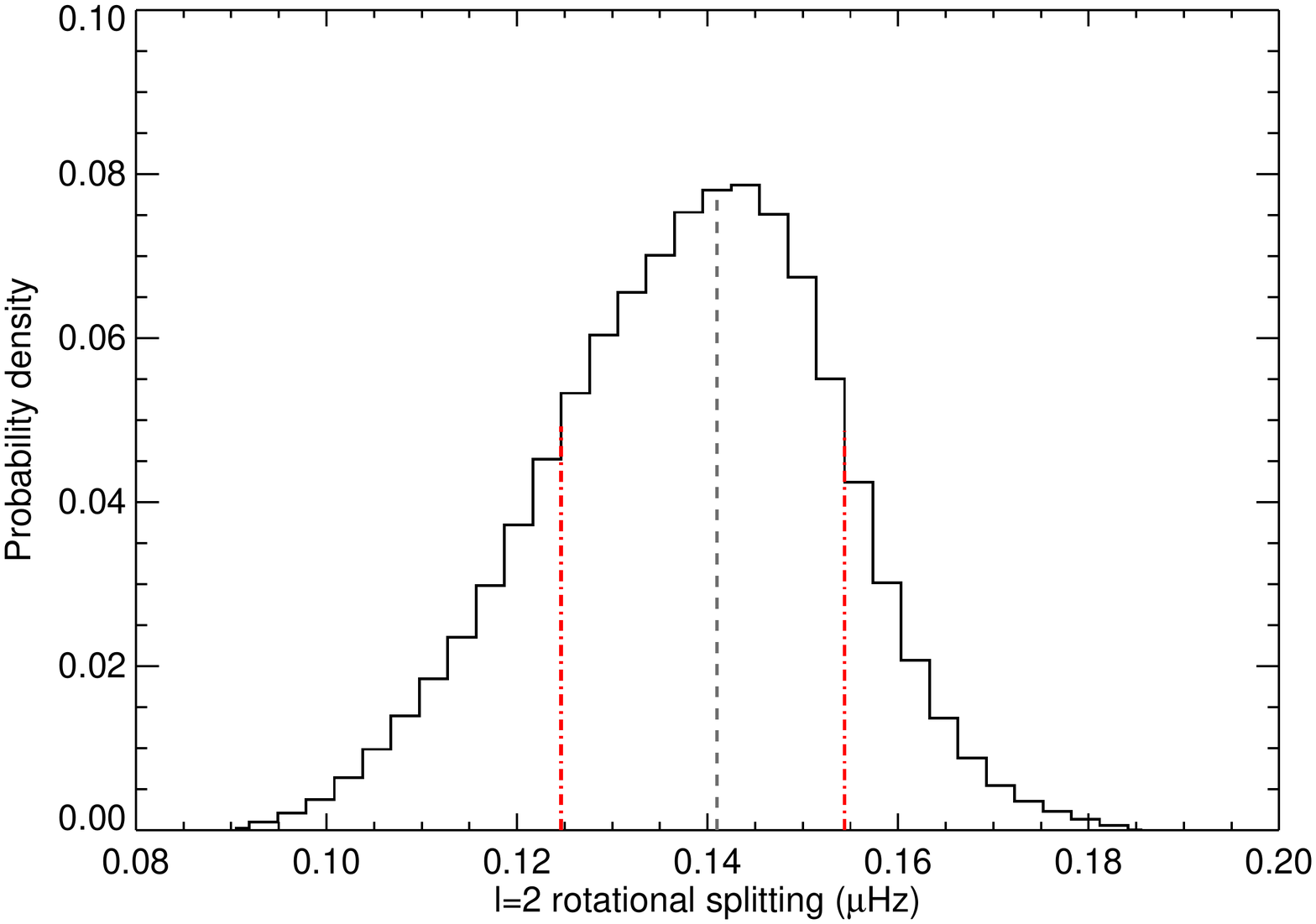}
	\caption{Probability density function for the averaged $\ell=2$ rotational splitting measured on the original spectrum (left) and on the modified spectrum (right) of KIC~9574283.  Same legend as in Fig.~\ref{fig_visibilityl2}.}
\label{Fig:Annex-2}
\end{center}
\end{figure*}
\begin{figure*}[!htbp]
\begin{center}
	\includegraphics[width=7cm,trim=8cm 2cm 2cm 0cm]{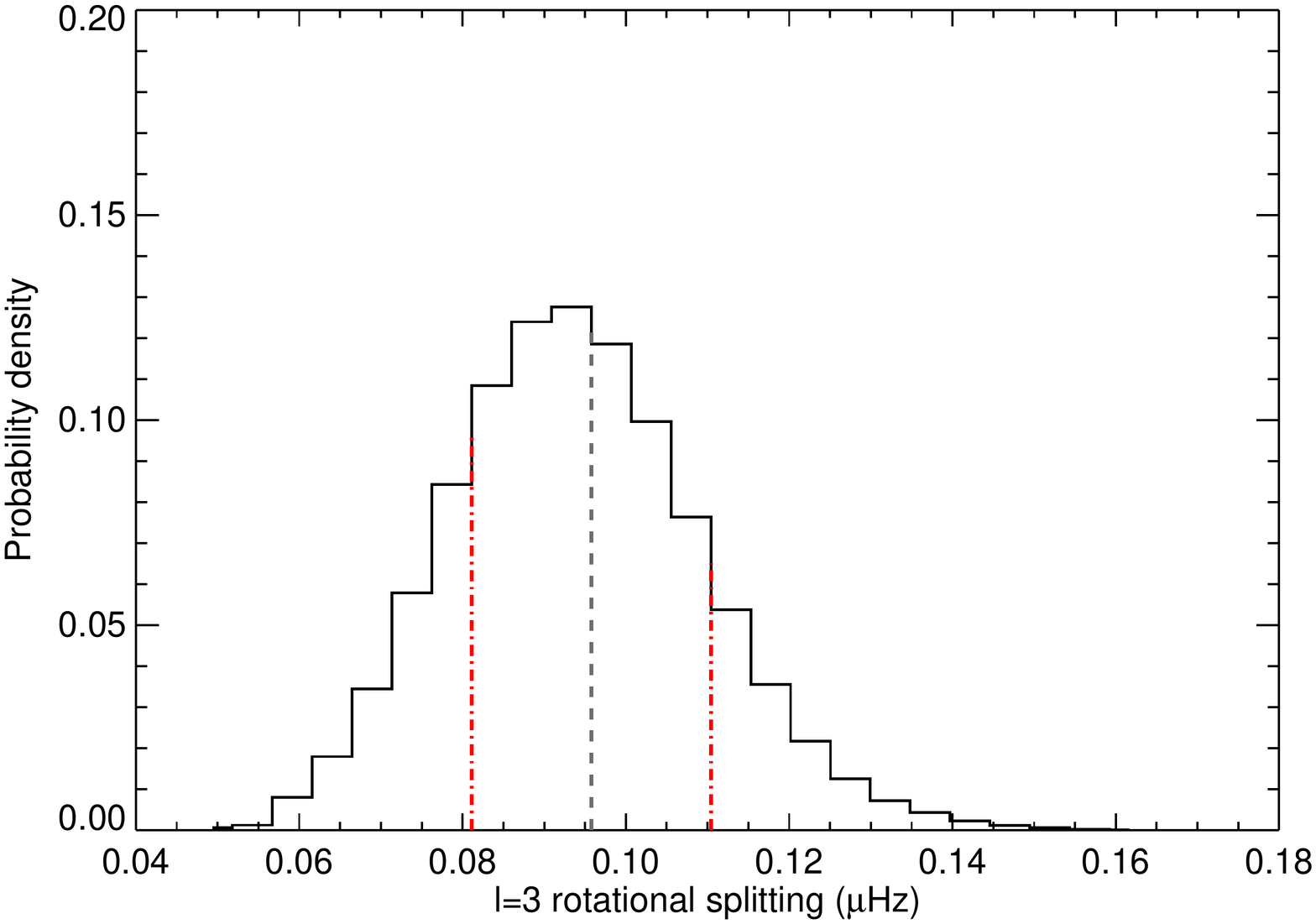}
	\includegraphics[width=7cm,trim=3cm 2cm 7cm 2cm]{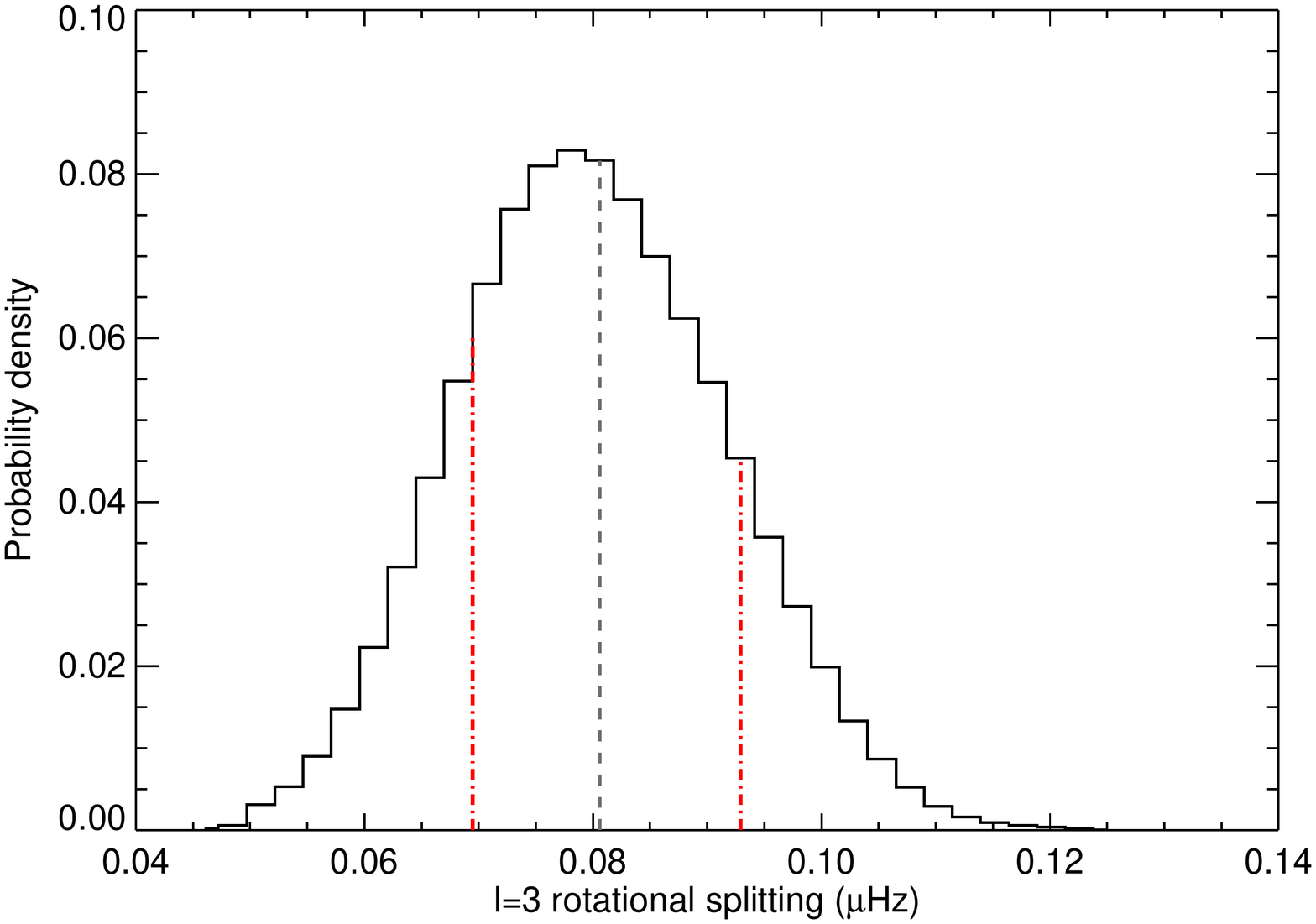}
	\caption{Same as Fig.\ref{Fig:Annex-2} but for $\ell=3$ averaged modes. Same legend as in Fig.~\ref{fig_visibilityl2}.}
\label{Fig:Annex-3}
\end{center}
\end{figure*}

We compared the measured mode parameters (frequencies, width, height, splitting) before and after modification of the reference spectrum. Although the altered spectrum provides larger uncertainties, results remain compatible at $1\sigma$ level. However, we notice that the uncertainty in the inclination angle is increased by the lack of information from the dipole modes, although the median remains quite robust (see Fig.\ref{Fig:Annex-2} and Fig.\ref{Fig:Annex-3}). 
We also notice that the dipole splittings measured on the altered spectrum can be measured at best with a precision of $\sigma=0.05$ $\mu$Hz (mode of SNR around 3) and at worst with a precision of $\sigma=0.6$ $\mu$Hz (mode of SNR around 1). 
Although the lower precision is typical of what we measured for $\ell=1$ in KIC~8561221, the best precision is much better than what is achievable in this star. Note also that $\ell=2$ and $\ell=3$ splittings of the reference star are changed because of the increase in the uncertainty of the inclination angle, even if we did not modified these mode properties. In the reference spectrum the inclination angle $i \approx 45^\circ$ means that $\ell=1$ are triplets of the same amplitude. This eases the measure of the splitting and explains the precision of $\sigma=0.05$ achieved in the best case.

Our tests therefore suggests that in the case of KIC~8561221, the inclination angle of $\sim20^\circ$ is probably correct. Indeed, if the inclination were higher (say $i>40$), the splitting of the dipole modes would have been well constrained, as well as the inclination. This is not what we obtained in the analysis of KIC~8561221, which indicates a low-inclination angle.

\end{appendix}

\bibliographystyle{aa}
\bibliography{/Users/rgarcia/Documents/BibReader/BIBLIO}

\end{document}